\documentclass{ieeetmlcn}

\usepackage[utf8]{inputenc}
\usepackage[T1]{fontenc}
\usepackage{amsmath,amssymb,amsfonts}
\usepackage{bm}
\usepackage{graphicx}
\usepackage{etoolbox}
\usepackage[hidelinks]{hyperref}
\usepackage[nocompress]{cite}
\usepackage{booktabs}
\usepackage{tabularx}
\usepackage{stfloats}
\usepackage{tikz}
\usetikzlibrary{positioning, arrows.meta, calc, fit, shapes.geometric, decorations.pathreplacing}
\usepackage{pgfplots}
\pgfplotsset{compat=1.18}
\usepgfplotslibrary{groupplots}
\usepackage{pgfplotstable}
\usepackage[caption=false]{subfig}
\usepackage[table]{xcolor}
\usepackage{booktabs}
\usepackage[acronym,nomain,nopostdot]{glossaries}
\glsdisablehyper

\usepackage[mode=text]{siunitx}
\DeclareSIUnit{\resourceelement}{RE}

\definecolor{myred}{HTML}{9E292B}
\definecolor{myblue}{HTML}{235787}
\definecolor{mygreen}{HTML}{5E6638}
\definecolor{mygray}{HTML}{444444}
\definecolor{myblack}{HTML}{000000}
\definecolor{mywhite}{HTML}{FFFFFF}
\definecolor{myyellow}{HTML}{BF892C}

\definecolor{myaltred}{HTML}{D46A78}
\definecolor{myaltblue}{HTML}{6699C2}
\definecolor{myaltgreen}{HTML}{B0B58C}
\definecolor{myaltgray}{HTML}{AAAAAA}

\definecolor{mydarkred}{HTML}{741E1A}
\definecolor{mylightred1}{HTML}{B15455}
\definecolor{mylightred2}{HTML}{C57F80}
\definecolor{mylightred3}{HTML}{D8A9AA}
\definecolor{mylightred4}{HTML}{ECD4D5}
\definecolor{mylightblue1}{HTML}{5A7DA5}
\definecolor{mylightblue2}{HTML}{7D99BA}
\definecolor{mylightblue3}{HTML}{B3C3D7}
\definecolor{mylightblue4}{HTML}{D3DCE8}
\definecolor{mydarkgreen}{HTML}{3E4822}
\definecolor{mylightgreen1}{HTML}{828859}
\definecolor{mylightgreen2}{HTML}{9AA075}
\definecolor{mylightgreen3}{HTML}{B8BC96}
\definecolor{mylightgreen4}{HTML}{D4D4B8}
\definecolor{mylightgray1}{HTML}{6F6F6F}
\definecolor{mylightgray2}{HTML}{999999}
\definecolor{mylightgray3}{HTML}{B4B4B4}
\definecolor{mylightgray4}{HTML}{DCDCDC}

\definecolor{nvidiagreen}{HTML}{76B900}
\definecolor{nvidiapurple}{HTML}{952FC6}
\definecolor{nvidiaorange}{HTML}{EF9100}
\definecolor{nvidiamagenta}{HTML}{D2308E}
\definecolor{nvidiateal}{HTML}{1DBBA4}
\definecolor{nvidiagray}{HTML}{757575}

\definecolor{greenlight2}{HTML}{CFFF40}
\definecolor{greenlight1}{HTML}{BFF230}
\definecolor{greendark1}{HTML}{3F8500}
\definecolor{greendark2}{HTML}{265600}

\definecolor{purplelight2}{HTML}{F9D4FF}
\definecolor{purplelight1}{HTML}{C359EF}
\definecolor{purpledark1}{HTML}{741D9D}
\definecolor{purpledark2}{HTML}{4D1368}

\definecolor{orangelight2}{HTML}{FEEEB2}
\definecolor{orangelight1}{HTML}{FCDE7B}
\definecolor{orangedark1}{HTML}{EF9100}
\definecolor{orangedark2}{HTML}{DF6500}

\definecolor{magentalight2}{HTML}{FFD3F2}
\definecolor{magentalight1}{HTML}{FC79CA}
\definecolor{magentadark1}{HTML}{8C1C55}
\definecolor{magentadark2}{HTML}{5D1337}

\definecolor{teallight2}{HTML}{ADFCF8}
\definecolor{teallight1}{HTML}{9AEFE5}
\definecolor{tealdark1}{HTML}{0D8473}
\definecolor{tealdark2}{HTML}{04554B}

\definecolor{mpltab0}{RGB}{31,119,180}
\definecolor{mpltab1}{RGB}{255,127,14}
\definecolor{mpltab2}{RGB}{44,160,44}
\definecolor{mpltab3}{RGB}{214,39,40}
\definecolor{mpltab4}{RGB}{148,103,189}
\definecolor{mpltab5}{RGB}{140,86,75}
\definecolor{mpltab6}{RGB}{227,119,194}
\definecolor{mpltab7}{RGB}{127,127,127}
\definecolor{mpltab8}{RGB}{188,189,34}
\definecolor{mpltab9}{RGB}{23,190,207}
\pgfplotscreateplotcyclelist{matplotlib}{
  {mpltab0}\\{mpltab1}\\{mpltab2}\\{mpltab3}\\{mpltab4}\\%
  {mpltab5}\\{mpltab6}\\{mpltab7}\\{mpltab8}\\{mpltab9}\\%
}

\tikzset{
  labelstyle/.style={
    fill=white, fill opacity=0.85, text opacity=1, inner sep=0.2mm, rotate=0, font=\footnotesize
  },
}

\pgfplotsset{
ber_plotstyle/.style={
width=6cm, 
height=3.5cm,
scale only axis,
axis background/.style={fill=none},
axis x line*=bottom,  
axis y line*=left,  
clip mode=individual,
every outer y axis line/.append style={black, very thick, cap=rect},
every outer x axis line/.append style={black, very thick, cap=rect},
ymax=1,
xmajorgrids,
ymajorgrids,
yminorgrids,
xlabel={SNR (\unit{\decibel})},
ylabel style = {black, font=\small, yshift=0.1cm},
xlabel style = {black, font=\small, yshift=-0.1cm},
yticklabel style = {font=\footnotesize},
xticklabel style = {font=\footnotesize, black},
tick align=outside,
tick style={black, very thick},
line width={2pt},
major tick length=2.25pt, 
minor tick length=0pt,
xtick pos=bottom,      
ytick pos=left,
}
}

\pgfplotsset{
standard_plotstyle/.style={
width=6cm, 
height=3.5cm,
scale only axis,
axis background/.style={fill=none},
axis x line*=bottom,  
axis y line*=left,  
clip mode=individual,
every outer y axis line/.append style={black, very thick, cap=rect},
every outer x axis line/.append style={black, very thick, cap=rect},
xmajorgrids,
ymajorgrids,
yminorgrids,
ylabel style = {black, font=\small, yshift=0.cm},
xlabel style = {black, font=\small, yshift=0.cm},
yticklabel style = {font=\footnotesize\rmfamily, text width=0.5cm, align=right},
xticklabel style = {font=\footnotesize, black},
tick align=outside,
tick style={black, very thick},
line width={2pt},
major tick length=2.25pt, 
minor tick length=0pt,
xtick pos=bottom,      
ytick pos=left,
}
}

\renewcommand{\vec}[1]{\mathbf{#1}}

\newcommand{\xv}{\vec{x}}
\newcommand{\yv}{\vec{y}}

\newcommand{\onev}{\vec{1}}

\newcommand{\Hm}{\vec{H}}

\newcommand{\Wm}{\vec{W}}
\newcommand{\Xm}{\vec{X}}
\newcommand{\Ym}{\vec{Y}}

\newcommand{\Cc}{{\cal C}}

\newcommand{\Ec}{{\cal E}}

\newcommand{\Gc}{{\cal G}}

\newcommand{\Mc}{{\cal M}}
\newcommand{\Nc}{{\cal N}}

\newcommand{\Pc}{{\cal P}}

\newcommand{\Rc}{{\cal R}}

\newcommand{\CN}{\mathcal{CN}}

\newcommand{\CC}{\mathbb{C}}

\newcommand{\RR}{\mathbb{R}}

\newcommand{\ZZ}{\mathbb{Z}}

\newcommand{\htp}{^{\mathsf{H}}}

\newcommand{\LB}{\left(}
\newcommand{\RB}{\right)}
\newcommand{\LP}{\left\{}
\newcommand{\RP}{\right\}}

\renewcommand{\ln}[1]{\mathop{\mathrm{ln}}\LB #1\RB}

\renewcommand{\exp}[1]{\mathop{\mathrm{exp}}\LB #1\RB}

\newcommand{\EE}{{\mathbb{E}}}

\newcommand\abs[1]{\left| #1 \right|}

\newcommand{\Df}{\Delta f}
\newcommand{\Ts}{T_{\mathrm s}}

\newcommand{\Tcp}{T_{\mathrm{cp}}}

\newcommand{\Dir}{\mathcal{D}}
\newcommand{\Wt}{\mathcal{W}_{\tau}}
\newcommand{\Wn}{\mathcal{W}_{\nu}}
\newcommand{\Wtt}{\widetilde{\mathcal{W}}_{\tau}}
\newcommand{\Wnt}{\widetilde{\mathcal{W}}_{\nu}}
\newcommand{\modM}[1]{\left(#1\right)_{M}}
\newcommand{\modN}[1]{\left(#1\right)_{N}}

\newcommand{\topk}{\ensuremath{\text{top-}k}}

\newcommand{\speedup}[2]{\pgfmathparse{#1/#2}\pgfmathprintnumber[fixed, zerofill, precision=1, assume math mode=true]{\pgfmathresult}$\times$}

\newacronym{otfs}{OTFS}{orthogonal time--frequency space}
\newacronym{cp}{CP}{cyclic prefix}
\newacronym{ofdm}{OFDM}{orthogonal frequency-division multiplexing}
\newacronym{dd}{DD}{delay--Doppler}
\newacronym{awgn}{AWGN}{additive white Gaussian noise}
\newacronym{dft}{DFT}{discrete Fourier transform}
\newacronym{qam}{QAM}{quadrature amplitude modulation}
\newacronym{ldpc}{LDPC}{low-density parity-check}
\newacronym{bler}{BLER}{block error rate}
\newacronym{nve}{NVE}{normalized validation error}
\newacronym{snr}{SNR}{signal-to-noise ratio}
\newacronym{ep}{EP}{expectation propagation}
\newacronym{csi}{CSI}{channel state information}
\newacronym{llr}{LLR}{log-likelihood ratio}
\newacronym{los}{LoS}{line-of-sight}
\newacronym{lmmse}{LMMSE}{linear minimum mean-square error}
\newacronym{mp}{MP}{message passing}
\newacronym{uamp}{UAMP}{unitary approximate message passing}
\newacronym{amp}{AMP}{approximate message passing}
\newacronym{gamp}{GAMP}{generalized approximate message passing}
\newacronym{vamp}{VAMP}{vector approximate message passing}
\newacronym{oamp}{OAMP}{orthogonal approximate message passing}
\newacronym{ec}{EC}{expectation consistency}
\newacronym{admm}{ADMM}{alternating direction method of multipliers}
\newacronym{bp}{BP}{belief propagation}
\newacronym{pcg}{PCG}{preconditioned conjugate gradient}
\newacronym{pic}{PIC}{parallel interference cancellation}
\newacronym{sic}{SIC}{successive interference cancellation}
\newacronym{llm}{LLM}{large language model}
\newacronym{gpu}{GPU}{graphics processing unit}
\newacronym{re}{RE}{resource element}
\newacronym{dmrs}{DMRS}{demodulation reference signal}
\newacronym{em}{EM}{expectation--maximization}
\newacronym{map}{MAP}{maximum a posteriori}
\newacronym{nlos}{NLoS}{non-line-of-sight}
\newacronym{ai}{AI}{artificial intelligence}
\newacronym{gp}{GP}{genetic programming}
\newacronym{aite}{AITE}{The AI Telco Engineer}
\newacronym{api}{API}{application programming interface}

\AtBeginDocument{%
    \definecolor{tmlcncolor}{cmyk}{0.93,0.59,0.15,0.02}%
    \definecolor{NavyBlue}{RGB}{0,86,125}%
}
\def\OJlogo{\vspace{-4pt}\includegraphics[height=18pt]{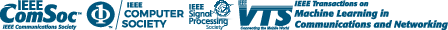}}
\def\seclogo{\vspace{10pt}\includegraphics[height=18pt]{new_logo}}

\def\OJlogo{}
\def\seclogo{}
\makeatletter
\patchcmd{\ps@plain}
  {This work is licensed under a Creative Commons Attribution 4.0 License. For more information, see https://creativecommons.org/licenses/by/4.0/}
  {}{}{}
\patchcmd{\ps@plain}
  {This work is licensed under a Creative Commons Attribution 4.0 License. For more information, see https://creativecommons.org/licenses/by/4.0/}
  {}{}{}
\makeatother

\def\authorrefmark#1{\ensuremath{^{\textbf{#1}}}}

\makeatletter
\newcommand{\tmlcn@publicationline}{%
    \ifdefempty{\@receiveddate}{%
        \ifdefempty{\@editor}{}{{\receivedfont\@editor\par}}%
    }{{\receivedfont Received\ \@receiveddate
        \ifdefempty{\@reviseddate}{}{; revised\ \@reviseddate}%
        \ifdefempty{\@accepteddate}{}{; accepted\ \@accepteddate}%
        \ifdefempty{\@publisheddate}{}{; Date of publication\nobreakspace\@publisheddate}%
        \ifdefempty{\@currentdate}{}{; date of current version\nobreakspace\@currentdate}%
        .\space\ifdefempty{\@editor}{}{\@editor}\par}}%
}
\patchcmd{\@maketitle}
    {{\receivedfont \ifx\@receiveddate\@empty\else Received\ \@receiveddate\fi\ifx\@revised\@empty\else;\ revised\ \@reviseddate\fi\ifx\@accepteddate\@empty\else;\ accepted\ \@accepteddate\fi\ifx\@publisheddate\@empty\else; Date of publication\nobreakspace\@publisheddate\fi\ifx\@currentdate\@empty\else;\space date of current version\nobreakspace\@currentdate\fi.\space\ifx\@editor\@empty\else\@editor\fi\par}}
    {\tmlcn@publicationline}
    {}
    {\ClassWarning{ieeetmlcn}{Could not patch the publication-metadata line}}

\newcommand{\tmlcn@volumeinfo}{%
    \ifdefempty{\@jvol}{%
        \ifdefempty{\@pubyear}{}{VOLUME\ \@jvol,\ \@pubyear}%
    }{VOLUME\ \@jvol,\ \@pubyear}%
}
\patchcmd{\ps@headings}
    {VOLUME\ \@jvol,\ \@pubyear}{\tmlcn@volumeinfo}{}
    {\ClassWarning{ieeetmlcn}{Could not patch the first running footer}}
\patchcmd{\ps@headings}
    {VOLUME\ \@jvol,\ \@pubyear}{\tmlcn@volumeinfo}{}
    {\ClassWarning{ieeetmlcn}{Could not patch the second running footer}}
\patchcmd{\ps@plain}
    {VOLUME\ \@jvol,\ \@pubyear}{\tmlcn@volumeinfo}{}
    {\ClassWarning{ieeetmlcn}{Could not patch the first title-page footer}}
\patchcmd{\ps@plain}
    {VOLUME\ \@jvol,\ \@pubyear}{\tmlcn@volumeinfo}{}
    {\ClassWarning{ieeetmlcn}{Could not patch the second title-page footer}}
\ps@headings
\makeatother

\title{Autonomous Discovery of Wireless Communications Algorithms}

\author{Fayçal Aït Aoudia\authorrefmark{1}, Member, IEEE,
Jakob~Hoydis\authorrefmark{1}, Fellow, IEEE,
Sebastian~Cammerer\authorrefmark{2}, Senior Member, IEEE,
Gian~Marti\authorrefmark{3}, Member, IEEE,\\
Merlin~Nimier-David\authorrefmark{3}, Nicolas Roussel\authorrefmark{3},
and Alexander~Keller\authorrefmark{2}}
\hypersetup{
    pdftitle={Autonomous Discovery of Wireless Communications Algorithms},
    pdfauthor={Fayçal Aït Aoudia; Jakob Hoydis; Sebastian Cammerer; Gian Marti; Merlin Nimier-David; Nicolas Roussel; Alexander Keller},
    pdfsubject={IEEE Transactions on Machine Learning in Communications and Networking},
    pdfkeywords={algorithm discovery, evolutionary search, large language model (LLM), pilotless communication, orthogonal frequency-time space (OTFS), wireless communications}
}
\affil{NVIDIA, 92400 Courbevoie, France}
\affil{NVIDIA, 10623 Berlin, Germany}
\affil{NVIDIA, 8004 Zürich, Switzerland}
\corresp{Corresponding author: Fayçal Aït Aoudia
(e-mail: \texttt{faitaoudia@nvidia.com}).}
\authornote{Parts of this article were presented at European Wireless,
June 2026~\cite{aite2026preprint}. Code repository:
\url{https://github.com/nvlabs/the-ai-telco-engineer}}

\begin{document}
\bstctlcite{IEEEtran:BSTcontrol}

\begin{abstract}
\Gls{llm}-driven evolutionary search is an emerging algorithm-discovery paradigm that has already produced novel results in several scientific fields. Yet its application to wireless communications remains largely unexplored. To bridge this gap, we introduce \gls{aite}, a framework to autonomously design algorithms for complex communication problems, while navigating performance--complexity tradeoffs. We showcase \gls{aite} on two challenging physical-layer problems: designing an equalizer for an \gls{otfs} system, and constructing a receiver algorithm for an \gls{ofdm} system using a custom constellation and operating without pilots. For the first task, \gls{aite} develops algorithms that outperform the best-known solutions while reducing computational latency by a factor of 3.6 compared to the strongest baseline. For the second task, it discovers the first explicit, explainable algorithms that achieve performance parity with state-of-the-art neural receivers. These results demonstrate the strong potential of \gls{llm}-driven evolutionary search for the autonomous discovery of next-generation wireless communications algorithms.
\end{abstract}


\maketitle
\glsresetall

\section{Introduction}

\IEEEPARstart{A}{lgorithms} are central to modern communication systems. They largely determine reliability
and efficiency, and serve as important product differentiators. Designing and
implementing such algorithms is a demanding task for communications system
engineers, requiring domain expertise, mathematical insight, and extensive
experimentation to identify robust implementations and suitable hyperparameters.
Moreover, solutions that perform well analytically or in simulation may behave
unexpectedly once integrated into a full system, where finding the root cause
may be difficult. \Glspl{llm} and agentic \gls{ai} have advanced rapidly, creating an opportunity to dramatically reduce the human effort required for algorithm development. In this article, we therefore investigate their ability to autonomously discover, implement, and optimize wireless communication algorithms.

An algorithm exists at two levels: an abstract mathematical description and a concrete implementation that can be executed. Thus, the optimization of an algorithm can be cast as a programming problem. The target of the problem depends on properties of the concrete implementation, such as its accuracy or latency when run on a specific hardware platform. One way of tackling such problems is \gls{gp}~\cite{koza1992genetic}. \Gls{gp} relies on evolutionary algorithms to improve computer programs by iteratively applying genetic operators, such as random mutation and crossover, to a generation of seed programs. One of the main problems of \gls{gp} relates to the definition of the search space, i.e., the set of valid code modifications, which makes scaling to complex programs difficult and requires deep domain expertise.

The integration of \glspl{llm} into \gls{gp} has gradually expanded the field into \gls{llm}-driven evolutionary search. The authors of~\cite{lehman2022evolution} were among the first to replace \gls{gp}'s mutation operator with an \gls{llm} that actually understands the code and provides logical program modifications, instead of random edits. This idea was further developed in~\cite{meyerson2024lmx} and~\cite{chen2023evoprompting}, where \glspl{llm} were also used for crossover operations. The first provably new scientific results were generated by FunSearch~\cite{romeraparedes2024funsearch}, which leveraged a fixed programmatic skeleton restricting the \gls{llm}'s search to an isolated target function, combined with island-based population management. This idea was scaled to multifile codebases in AlphaEvolve~\cite{novikov2025alphaevolve}, which was reproduced shortly thereafter as the open-source framework OpenEvolve~\cite{openevolve}. Since then, there has been a rapid proliferation of \gls{llm}-based evolutionary methods for algorithm discovery, such as ShinkaEvolve~\cite{lange2025shinka}, GEPA~\cite{agrawal2025gepa}, and the SkyDiscover framework~\cite{skydiscover} (including AdaEvolve~\cite{cemri2026adaevolve} and EvoX~\cite{liu2026evox}). They introduce several critical capabilities, such as embedding-based novelty rejection, multiarmed bandit \gls{llm} selection, adaptive exploration and compute budget control, as well as optimization of the search strategy code itself. Recently, the code-generation step was extended from a single \gls{llm} invocation to tool-using coding agents that iteratively implement, evaluate, and refine solutions before submitting them~\cite{chen2026avo,aite2026preprint}. This extension is particularly important for challenging coding problems that require the use of debugging and profiling tools or documentation lookups.

In this article, we report on initial experiments applying \gls{llm}-driven evolutionary search to the design of wireless communications algorithms, providing a useful starting point for others interested in this direction. Notably, even these early results match or surpass the state of the art on the considered tasks, suggesting that agentic \gls{ai} may have crossed a critical capability threshold, with profound implications for the future of research and development in our field.

The only directly related publications we are aware of are~\cite{weindel2026deletion} and~\cite{autoport}, which apply \gls{llm}-driven evolutionary search to the optimization of deletion-correcting codes and antenna port selection in a fluid antenna system, respectively. The authors of~\cite{genesis} describe a versatile platform capable of automating research within a large-scale radio testbed, without providing detailed technical results.

The rest of this article is structured as follows. Section~\ref{sec:architecture} describes the architecture of our evolutionary search framework called \gls{aite}, which is available as open-source software~\cite{aite2026repo} and implemented in Python using LangChain~\cite{langchain}. We did not rely on existing frameworks such as~\cite{skydiscover, lange2025shinka, openevolve}, although these could have been adapted.
In Sections~\ref{sec:otfs-equalizer-design} and~\ref{sec:pilotless}, we describe two specific physical-layer problems that \gls{aite} was tasked with, namely developing an equalizer for an \gls{otfs} system~\cite{otfsbits1} as well as a receiver algorithm for an \gls{ofdm} system operating without pilots and with custom constellations~\cite{9508784}. We choose these two tasks because, although both are nontrivial, they are very different in nature. Equalizer design for \gls{otfs} is well studied, so the main challenge is to find efficient implementations that simultaneously improve on existing solutions. The pilotless receiver, in contrast, has essentially no prior art, so the biggest challenge is to discover a working algorithm.
For the first task, \gls{aite} found algorithms that outperform state-of-the-art equalizers while running about $3.6\times$ faster. For the second task, only solutions based on neural networks were known, and \gls{aite} found novel explicit algorithms with comparable performance.
Section~\ref{sec:conclusion} concludes the article with a discussion of open problems and future directions. The Appendices provide detailed descriptions of selected algorithms generated by \gls{aite} and illustrate the depth and implementation complexity of the discovered solutions.

\section{Architecture of The AI Telco Engineer}
\label{sec:architecture}

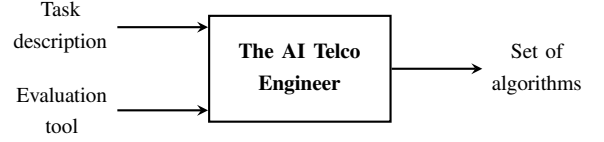
\begin{figure}[t]
    \centering
    \begin{tikzpicture}[
    box/.style={thick, draw, minimum width=2.4cm, minimum height=1.4cm, align=center, font=\footnotesize\bfseries},
    lbl/.style={font=\footnotesize, align=center},
    arr/.style={thick, ->, >=stealth}
]
    \node[box] (fw) {The AI Telco\\Engineer};

    \node[lbl, left=1.2cm of fw, yshift=0.55cm]  (task) {Task\\description};
    \node[lbl, left=1.2cm of fw, yshift=-0.55cm] (eval) {Evaluation\\tool};

    \node[lbl, right=1.2cm of fw] (out) {Set of\\algorithms};

    \draw[arr] (task.east) -- (task.east -| fw.west);
    \draw[arr] (eval.east) -- (eval.east -| fw.west);
    \draw[arr] (fw.east) -- (out.west);
\end{tikzpicture}
    \caption{\gls{aite} from an input--output perspective. The user describes a task, i.e., a problem description paired with an evaluation tool, and \gls{aite} returns a set of algorithms that explores the metric--complexity tradeoff.}
    \label{fig:io_view}
\end{figure}

\Gls{aite} is an \gls{llm}-driven framework that iteratively explores the space of algorithms that solve a specified problem. An input--output view of the system is shown in Fig.~\ref{fig:io_view}. \gls{aite} takes as input a \emph{task}, defined by a natural-language problem description paired with an immutable, task-specific \emph{evaluation tool}. The evaluation tool executes algorithm implementations and returns a scalar metric evaluating the algorithm's performance and a scalar measuring the complexity of the implementation (e.g., via the measured latency). Because the evaluation tool is not editable by the system and can only be invoked, \gls{aite} prevents ``metric gaming'' in which the scoring procedure is modified to artificially inflate performance.

\Gls{aite} does not focus on evolving a single best solution, but rather aims to output a \emph{set} of solutions that spans the performance--complexity tradeoff. Concretely, solutions are positioned in the two-dimensional plane spanned by the task metric and the complexity measure returned by the evaluation tool, yielding a point cloud from which the user can pick an operating point after the run, without being involved in the optimization itself. The optimization objective is to expand the Pareto front by iteratively evolving a population of algorithms. The same principle would extend naturally to a higher-dimensional objective space, although we have not explored this in our experiments.

\subsection{Global Architecture}
\label{subsec:global_architecture}
\begin{figure}[t]
    \centering
    \resizebox{\columnwidth}{!}{\begin{tikzpicture}[
    box/.style={draw, thick, minimum height=0.8cm, align=center, font=\footnotesize},
    orch/.style={box, thick, minimum width=2.2cm, font=\footnotesize\bfseries},
    idea/.style={minimum height=0.8cm, minimum width=1.0cm, align=center, font=\footnotesize},
    worker/.style={draw, thick, minimum width=0.9cm, minimum height=0.9cm, align=center, font=\scriptsize},
    arr/.style={->, thick, >=stealth},
    lbl/.style={font=\scriptsize}
]
    \node[orch] (orch) {Orchestrator};

    \node[idea, below left=0.7cm and 1.2cm of orch]  (i1) {Idea 1};
    \node[idea, below=0.7cm of orch]                  (i2) {Idea 2};
    \node[idea, below right=0.7cm and 1.2cm of orch]  (in) {Idea $N$};

    \node at ($(i2)!0.5!(in)$) {\footnotesize$\cdots$};

    \node[worker, below=0.5cm of i1, xshift=-0.55cm] (a1) {Worker\\1};
    \node[worker, below=0.5cm of i1, xshift= 0.55cm] (a2) {Worker\\2};

    \node[worker, below=0.5cm of i2, xshift=-0.55cm] (a3) {Worker\\3};
    \node[worker, below=0.5cm of i2, xshift= 0.55cm] (a4) {Worker\\4};
    
    \node[worker, below=0.5cm of in, xshift=-0.55cm] (an1) {Worker\\$M\!-\!1$};
    \node[worker, below=0.5cm of in, xshift= 0.55cm] (an2) {Worker\\$M$};

    \node at ($(a4)!0.5!(an1)$) {\footnotesize$\cdots$};

    \node[draw, thick, dashed, rounded corners=3pt, inner sep=8pt,
          fit=(a1)(a2)(a3)(a4)(an1)(an2),
          label={[font=\footnotesize\bfseries, align=center]right:Worker\\pool}] (pool) {};

    \draw[arr] (orch.south) -- ([xshift=0.3cm, yshift=-0.05cm]i1.north);
    \draw[arr] (orch.south) -- (i2.north);
    \draw[arr] (orch.south) -- ([xshift=-0.3cm, yshift=-0.05cm]in.north);

    \draw[arr] (i1.south) -- (a1.north);
    \draw[arr] (i1.south) -- (a2.north);
    \draw[arr] (i2.south) -- (a3.north);
    \draw[arr] (i2.south) -- (a4.north);
    \draw[arr] (in.south) -- (an1.north);
    \draw[arr] (in.south) -- (an2.north);

    \draw[thick, decorate, decoration={brace, mirror, amplitude=5pt}]
        ([yshift=-0.15cm]pool.south west) -- ([yshift=-0.15cm]pool.south east)
        coordinate[midway, below=10pt] (brace-mid);

    \draw[arr, rounded corners=4pt]
        (brace-mid) -- ++(0, -0.3)
        -| ([xshift=-0.4cm]pool.west |- orch.west)
        -- (orch.west)
        node[lbl, above, pos=0.45] {Implemented algorithms};
\end{tikzpicture}}
    \caption{Overview of the idea-driven iterative optimization loop. The orchestrator generates ideas that are distributed to parallel workers, each operating in its own workspace. Implemented algorithms flow back to the orchestrator to seed the next generation.}
    \label{fig:overview}
\end{figure}
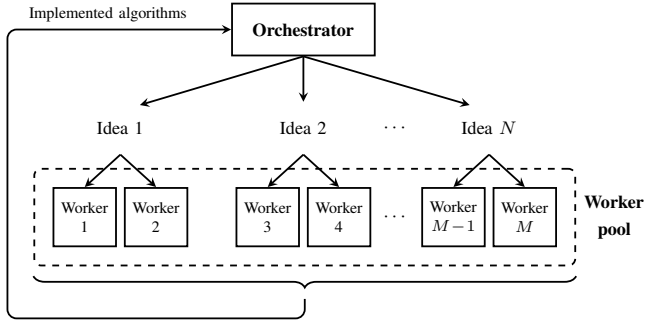

As illustrated in Fig.~\ref{fig:overview}, \gls{aite} follows a two-tier architecture inspired by~\cite{romeraparedes2024funsearch,novikov2025alphaevolve,lange2025shinka}.
Most importantly, it relies on agentic code generation as in~\cite{chen2026avo,aite2026preprint}, instead of a single \gls{llm} invocation. 
This architecture proved effective for the tasks considered here.
An \emph{orchestrator} drives the global optimization loop, while a pool of \emph{workers} implements and refines solutions in parallel. Optimization proceeds over multiple generations. At each generation, the orchestrator proposes \(N\) distinct abstract algorithmic ideas, and a population of \(M\) workers is distributed across these ideas, with \(M/N\) workers assigned to each. Each worker runs independently, in parallel, and within its own containerized environment referred to as a \emph{workspace}. Neither the orchestrator nor the workers have internet access.

Assigning multiple workers to the same idea exploits \gls{llm} stochasticity (assuming a nonzero decoding temperature). Different workers implementing the same idea tend to produce different implementations, thereby reducing the risk of discarding a promising idea due to a single poor implementation. After all workers in a generation complete, the orchestrator aggregates the resulting implementations, updates a global \emph{leaderboard}, and uses the accumulated outcomes to propose the next generation of ideas. Subsequent ideas can be entirely novel approaches, refinements of previously explored ones, or combinations thereof, depending on the current state of the performance--complexity frontier.

\subsection{Orchestrator}
\label{subsec:orchestrator}

\begin{figure*}[t]
    \centering
    \resizebox{\textwidth}{!}{\begin{tikzpicture}[
    x=1cm, y=1cm,
    box/.style={
        draw, thick,
        minimum height=0.6cm, minimum width=2.0cm,
        align=center, font=\fontsize{5.62}{7}\selectfont,
    },
    arr/.style={->, thick, >=stealth},
    lbl/.style={
        font=\fontsize{4.91}{6}\selectfont, align=center,
        fill=white, fill opacity=0.92, text opacity=1,
        inner sep=1.5pt,
    },
]
    \node[box] (ideas)    at (3.2, 0)     {Idea generation};
    \node[box] (lb)       at (9.4, 0)     {Leaderboard};
    \node[box] (summ)     at (6.5, -1.7)  {Post-processing};
    \node[box] (launcher) at (3.2, -1.7)  {Job launching};
    \node[box] (pref)     at (-1.3, -1.7) {Prompt refinement};

    \node[box, font=\fontsize{5.62}{7}\selectfont\bfseries] (pool) at (3.2, -3.3) {Worker pool};

    \node[draw, thick, dashed, rounded corners=3pt, inner sep=10pt,
          fit=(ideas)(lb)(summ)(launcher)(pref),
          label={[font=\fontsize{5.62}{7}\selectfont\bfseries]above:Orchestrator}] {};

    \draw[arr] (1.0, 1.05) -- (1.0, 0) -- (ideas.west);
    \draw[arr] (1.0, 0) -- (1.0, -0.85) -- (2.5, -0.85) -- (2.5, -1.4);
    \node[lbl] at (1.0, 1.25) {Task description};

    \draw[arr] ([yshift=3pt]ideas.east) -- ([yshift=3pt]lb.west)
        node[lbl, above=1pt, midway] {ideas};
    \draw[arr] ([yshift=-3pt]lb.west) -- ([yshift=-3pt]ideas.east)
        node[lbl, below=1pt, midway] {Pareto and\\off-front algorithms};

    \draw[arr] (ideas.south) -- (launcher.north);
    \node[lbl, anchor=east] at (3.15, -0.65) {ideas};

    \draw[arr] ([xshift=0.7cm]ideas.south) -- (3.9, -1.05) -- (6.5, -1.05) -- (summ.north);
    \node[lbl, above=1pt] at (5.2, -1.05) {ideas};

    \draw[arr] (launcher.south) -- (pool.north);
    \node[lbl, anchor=west] at (3.35, -2.5) {jobs};

    \draw[arr] ([yshift=0.2cm]pool.east) -- (6.5, -3.1) -- (summ.south);
    \node[lbl, anchor=west] at (6.65, -2.7) {code};

    \draw[arr] (summ.east) -- (8.7, -1.7) -- (8.7, -0.3);
    \node[lbl, anchor=west] at (8.85, -1.05) {summaries,\\verdicts};

    \draw[arr] ([yshift=-0.25cm]pool.east) -- (10.15, -3.55) -- (10.15, -0.3);
    \node[lbl, anchor=west] at (10.30, -3.2) {code, metric,\\complexity};

    \draw[arr] ([yshift=0.2cm]pool.west) -- (-1.3, -3.1) -- (pref.south);
    \node[lbl, above=1pt] at (-0.2, -3.1) {journals};

    \draw[arr] (0.7, -3.5) -- ([yshift=-0.20cm]pool.west);
    \node[lbl, anchor=east] at (0.65, -3.5) {Evaluation tool};

    \draw[arr] (pref.east) -- (launcher.west);
    \node[lbl, above=0.5pt] at (0.95, -1.7) {refined\\prompt};
\end{tikzpicture}}
    \caption{Orchestrator internal structure. Idea generation is conditioned on the task description together with Pareto-front and off-front algorithms drawn from the leaderboard. The job-launching stage dispatches ideas to the worker pool. Workers return code, metric, and complexity to the leaderboard. The post-processing module pairs each implementation with its assigned idea to produce summaries and adherence verdicts, which are also stored in the leaderboard. Worker journals feed the prompt-refinement stage, whose refined prompts are injected into subsequent jobs.}
    \label{fig:orchestrator_overview}
\end{figure*}
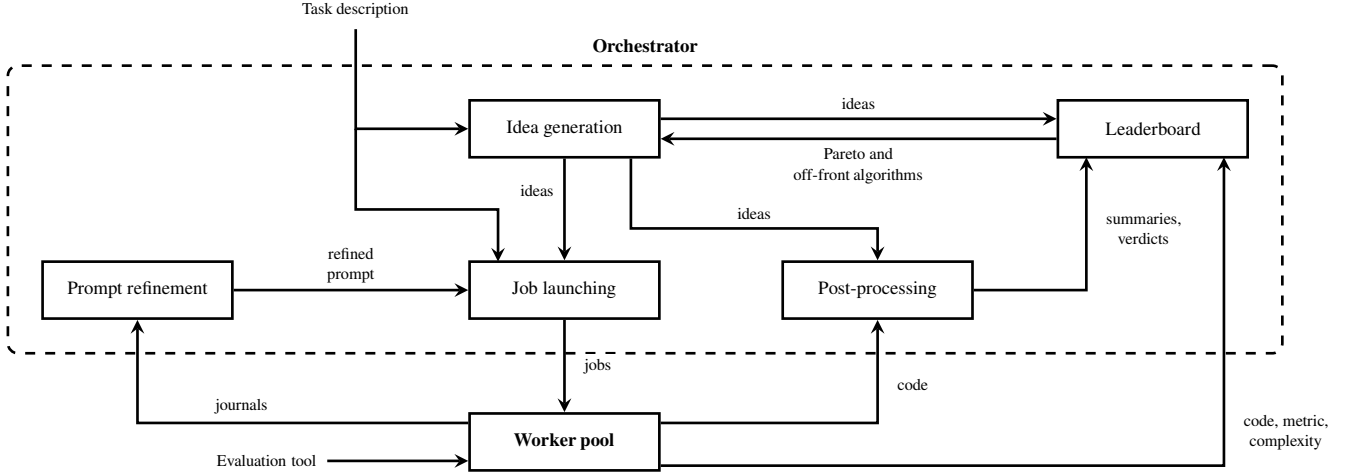

The orchestrator architecture is shown in Fig.~\ref{fig:orchestrator_overview}. It coordinates the optimization by (i) generating diverse ideas, (ii) dispatching them to workers, (iii) post-processing and summarizing the algorithms produced by the workers, and (iv) maintaining a global leaderboard that records all algorithm implementations and their evaluation scores. The leaderboard groups algorithms by their assigned idea and stores multiple (metric, complexity) evaluation points per algorithm when postrun hyperparameter tuning is enabled (Section~\ref{subsubsec:hp_finetuning}). The Pareto front across all implementations is also maintained.

\subsubsection{Idea Generation}
\label{subsubsec:idea_generation}

At each generation, the orchestrator \gls{llm} is instructed to produce \(N\) \emph{distinct} algorithmic ideas that span the metric--complexity tradeoff. Centralizing this step avoids a scenario where independent workers each choose their own approach and then tend to converge on similar methods, causing the explored space to collapse. Beyond the first generation, the idea budget must be split between \emph{refining} previously explored approaches (exploitation) and proposing \emph{fresh} ones (exploration). Because the number of workers, and hence the per-generation budget, is limited, this allocation is critical to search efficiency~\cite{lange2025shinka,cemri2026adaevolve}. \gls{aite} delegates this tradeoff to the orchestrator \gls{llm} and instructs it to consistently reserve a part of the budget for fresh ideas throughout the run.

For the initial generation, the orchestrator \gls{llm} receives only the user query and is instructed to output exactly \(N\) distinct algorithmic ideas, since there are no prior ones to refine. For subsequent generations, idea generation is conditioned on previously explored ideas and their corresponding implemented algorithms. However, providing the entire leaderboard to the orchestrator is not scalable due to the limited context window of the \gls{llm}.
Instead, \gls{aite} uses a two-part context construction: Pareto-front entries are always included, and additional off-front entries are sampled randomly. Off-front sampling includes at most one representative algorithm per idea and excludes algorithms that did not follow the assigned approach, based on a verdict produced by the post-processing stage (cf. Section~\ref{subsubsec:post_processing}).

Off-front entries are sampled without replacement using a temperature-controlled softmax over a spread-normalized metric gap. Let \(m_i\) denote the metric of algorithm \(i\) in the off-front pool and let \(m^\star\) be the best metric in the pool. Define the gap \(g_i = |m_i - m^\star|\) and the pool spread \(S = \max_i g_i\). The normalized gap is \(\tilde g_i = g_i / S\). If \(S=0\), we set \(\tilde g_i=0\) for all \(i\), resulting in uniform sampling. The sampling probability is then $p_i \propto \exp{-\tilde g_i / T}$,
where \(T>0\) is a temperature hyperparameter that remains fixed throughout the run. Smaller values of \(T\) yield greedier selection, whereas larger values yield nearly uniform sampling.

\subsubsection{Job Launching}
\label{subsubsec:job_launching}

Once a generation's ideas have been generated, the job-launching stage turns them into work units and dispatches them to the worker pool. The \(N\) ideas are distributed across the \(M\) available workers, with \(M/N\) workers assigned to each idea (see Fig.~\ref{fig:overview}). Every worker therefore receives a single idea and is tasked with implementing it in its own dedicated workspace.

For each worker, the orchestrator assembles a \emph{job}: a self-contained initial context that fully specifies what the worker must do. Beyond the assigned idea, this context includes the task description shared by all workers and, when refining ideas, references to one or more earlier implementations that the worker should build upon. Concretely, a job is materialized as a single prompt template into which the task description, the assigned idea, and the source code of optional reference algorithms are injected. The references' achieved metrics are deliberately withheld. Exposing them tends to bias the worker toward small edits of an existing solution rather than implementing the assigned approach.

As detailed in Section~\ref{subsubsec:prompt_refinement}, the prompt template is not fixed but refined across generations to improve worker efficiency, while protected placeholders guarantee that the task description and the assigned idea are preserved verbatim.

\subsubsection{Post-Processing of Generated Algorithms}
\label{subsubsec:post_processing}

Worker outputs do not always follow their assigned ideas, especially when an idea is difficult to implement or when the worker \gls{llm} fails to adhere to the provided approach description. We empirically observed that state-of-the-art \glspl{llm} (e.g., GPT-5.5) diverge from the assigned idea less frequently than smaller open-weight models (e.g., MiniMax-M2.5 or gpt-oss-120b). Regardless of its source, such divergence can mislead the orchestrator, e.g., into discarding an idea that appears to perform poorly when, in fact, it was never actually implemented.
To mitigate this, \gls{aite} post-processes each completed workspace by reading the resulting code and prompting the orchestrator \gls{llm} to produce a concise summary prefixed with a ternary verdict---\emph{yes}, \emph{partial}, or \emph{no}---indicating whether the implemented algorithm follows the assigned idea, independent of its achieved score.

These summaries serve two purposes. First, they provide a compact, human-readable description of what each algorithm actually implements, allowing the orchestrator to reason over a large history of explored methods. Second, they act as a filter when sampling off-front entries for idea generation (see Section~\ref{subsubsec:idea_generation}). Only the best-scoring implementations with a \emph{yes} verdict are eligible to represent an idea, preventing the orchestrator from being misled by spurious deviations.

\subsubsection{Worker Prompt Refinement}
\label{subsubsec:prompt_refinement}

\gls{aite} is designed to operate with different \glspl{llm}, which can vary widely in their tool-use patterns and ability to adhere to instructions. To improve worker efficiency across generations, \gls{aite} refines the worker prompt template using a dedicated prompt-refinement stage. After each worker completes, its execution journal (which records all \gls{llm} and tool calls together with their outcomes) is read and structured process statistics (e.g., number of evaluation attempts, file reads and writes, and timeouts) are extracted. An \gls{llm} is then used to produce a behavioral critique.

At the end of a generation, the orchestrator aggregates these behavioral critiques and requests an \gls{llm} to produce a refined prompt template for the next generation, which is used when launching the next generation's jobs (see Section~\ref{subsubsec:job_launching}). For example, if multiple workers make the same coding error, the refined prompt may include a warning to avoid this error in the future. The refinement affects only the generic instructions that govern how a worker operates. The task description and the assigned approach are injected through placeholders and are therefore never altered. This design lets \gls{aite} adapt to different \gls{llm} behaviors while keeping algorithm selection and exploration decisions centralized in the orchestrator's idea-generation stage.

\subsection{Worker Workflow}
\label{subsec:worker_workflow}

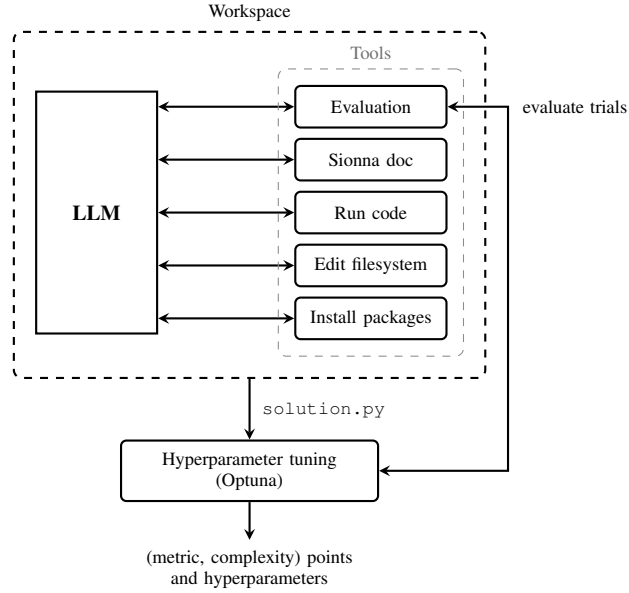
\begin{figure}[t]
    \centering
    \begin{tikzpicture}[
    agent/.style={thick, draw, minimum width=1.6cm, minimum height=3.2cm, align=center, font=\footnotesize\bfseries},
    tool/.style={thick, draw, rounded corners=2pt, minimum width=2cm, minimum height=0.55cm, align=center, font=\scriptsize},
    arr/.style={thick, <->, >=stealth},
    dir/.style={thick, ->, >=stealth},
]
    \node[agent] (agent) {LLM};

    \node[tool, right=1.8cm of agent, yshift=1.4cm]  (t1) {Evaluation};
    \node[tool, right=1.8cm of agent, yshift=0.7cm]  (t2) {Sionna doc};
    \node[tool, right=1.8cm of agent]                 (t3) {Run code};
    \node[tool, right=1.8cm of agent, yshift=-0.7cm] (t4) {Edit filesystem};
    \node[tool, right=1.8cm of agent, yshift=-1.4cm] (t5) {Install packages};

    \draw[arr] (agent.east |- t1) -- (t1.west);
    \draw[arr] (agent.east |- t2) -- (t2.west);
    \draw[arr] (agent.east |- t3) -- (t3.west);
    \draw[arr] (agent.east |- t4) -- (t4.west);
    \draw[arr] (agent.east |- t5) -- (t5.west);

    \coordinate (toolslabel) at ([yshift=0.42cm]t1.north);
    \node[draw=black!45, thin, dashed, rounded corners=3pt, inner sep=6pt,
          fit=(t1)(t5)] (tools) {};
    \node[font=\scriptsize, text=black!55] at (toolslabel) {Tools};

    \node[draw, thick, dashed, rounded corners=3pt, inner sep=8pt,
          fit=(agent)(tools)(toolslabel),
          label={[font=\scriptsize]above:Workspace}] (ws) {};

    \node[draw, thick, rounded corners=2pt, align=center, font=\scriptsize,
          minimum width=3.4cm, minimum height=0.8cm,
          below=0.8cm of ws] (hp) {Hyperparameter tuning\\(Optuna)};

    \draw[dir] (ws.south) -- (hp.north)
        node[midway, right=1pt, font=\scriptsize] {\texttt{solution.py}};

    \coordinate (riser) at ([xshift=0.8cm]t1.east);
    \draw[arr] (hp.east) -| (riser) -- (t1.east);
    \node[font=\scriptsize, anchor=west] at ([xshift=2pt]riser) {evaluate trials};

    \draw[dir] (hp.south) -- ++(0,-0.5)
        node[below, font=\scriptsize, align=center]
        {(metric, complexity) points\\and hyperparameters};
\end{tikzpicture}
    \caption{Worker architecture. Each worker runs an agentic ReAct loop~\cite{yao2023react} operating inside an isolated containerized workspace with access to tools, including the evaluation tool. Once the worker completes, an optional postrun stage tunes the algorithm's hyperparameters.}
    \label{fig:worker_workflow}
\end{figure}

Having described the orchestrator, we now turn to the workers that implement the ideas it dispatches. Workers operate in parallel, each running as a separate process that hosts a single \gls{llm}-driven ReAct agent~\cite{yao2023react} which repeatedly reasons about what to do, takes an action using a tool, observes the result, and adapts until it completes the task. The resulting worker architecture is shown in Fig.~\ref{fig:worker_workflow}. Tasks are distributed by a worker pool, which submits them to the worker processes and collects their completion results asynchronously. Each task is associated with a fresh workspace, isolated through a container interface to prevent interference across workers and to protect the host environment.

Within a workspace, the worker follows a two-file workflow: \texttt{draft.py} is used for iterative development and experimentation, while \texttt{solution.py} stores the best-performing code encountered so far.\footnote{Although we use Python in this work, having \gls{aite} write code in a different language is straightforward.}
The evaluation tool is invoked repeatedly by the worker \gls{llm} on \texttt{draft.py}. When a new best metric is achieved, the framework automatically copies \texttt{draft.py} to \texttt{solution.py}, ensuring that progress is preserved even if the worker is interrupted. The worker is terminated either when the worker \gls{llm} decides to stop or when a generation-level timeout is reached. In either case, \texttt{solution.py} is retained as the worker's output for downstream processing.

\subsubsection{Hyperparameter Fine-Tuning}
\label{subsubsec:hp_finetuning}

To reduce the risk of discarding a promising algorithm due to poorly chosen hyperparameters, \gls{aite} performs optional postrun hyperparameter tuning on each successful worker output, as shown in Fig.~\ref{fig:worker_workflow}. The worker \gls{llm} is instructed to declare tunable hyperparameters by calling one of the following forms:
{\small
\begin{verbatim}
HP.get("name", default, low=..., high=...)
HP.get("name", default, choices=[...])
\end{verbatim}
}
corresponding, respectively, to a bounded numerical range and a categorical choice set. The framework parses the abstract syntax tree of \texttt{solution.py} to extract the implied search space, then runs Bayesian multiobjective optimization using Optuna~\cite{optuna} over the task metric and complexity.

Each Optuna trial evaluates \texttt{solution.py} for one candidate hyperparameter configuration using the evaluation tool and yields a single (metric, complexity) point. Because the two objectives often conflict, tuning does not return a single best configuration but a set of Pareto-optimal ones. Each retained configuration is then attached to the worker's generated algorithm in the global leaderboard as a distinct (metric, complexity) point, tagged with its hyperparameter values. A single worker output may therefore contribute several leaderboard points, rather than just one.

\subsubsection{Sionna}
\label{subsubsec:sionna_doc}

In our experiments, \gls{aite} uses Sionna~\cite{sionna} to perform link-level simulations. To help workers write correct Sionna code (in PyTorch~\cite{ansel2024pytorch}), the framework provides a documentation tool offering three operations: \emph{Search}, \emph{Help}, and \emph{List}. \emph{Search} performs semantic retrieval over the Sionna \gls{api} docstrings and curated tutorials using a FAISS~\cite{johnson2021faiss} vector store with an external reranker. \emph{Help} retrieves the full docstring and signature of a Sionna symbol, while \emph{List} enumerates available classes and functions within a given Sionna module. 

\subsubsection{Task-Specific Evaluation Tool}
\label{subsubsec:evaluation_tool}

The evaluation tool is a critical component of the system. It is supplied by the user as part of the task specification and, by defining the metric and the complexity proxy to optimize, fixes the two objectives that drive the entire search. In our experience, it is also typically the bottleneck of the optimization, consuming the largest share of the compute and time budget. Communication algorithms are usually assessed through Monte Carlo simulations, which trade accuracy for compute. More accurate or exhaustive simulations therefore come at the cost of slower evaluations and, for a fixed time and compute budget, fewer generations of the framework.

In \gls{aite}, each task provides its evaluation tool as a pluggable component that executes worker code and returns a standardized, parseable output. The first line follows one of two forms: \texttt{SUCCESS, <metric>, <complexity>} or a bare \texttt{FAILURE}. A binary success/failure flag is thus always available and, on success, the returned metric and complexity define the algorithm's position in the two-objective plane. Beyond scoring, the evaluation tool can enforce constraints by returning a failure whenever a requirement is violated (e.g., when a secondary metric exceeds a threshold), thereby restricting metric optimization to a feasible set. It can also return error messages and warnings. To prevent any single evaluation from stalling the optimization loop, each invocation is subject to a configurable timeout. Moreover, workers can invoke the evaluation tool but cannot modify it, which prevents direct modification of the scoring procedure while still allowing arbitrary user-defined evaluation code.

\section{OTFS Equalizer Design}
\label{sec:otfs-equalizer-design}

The first task assigned to \gls{aite} is to implement an equalizer for an \gls{otfs} system~\cite{otfsbits1}.
In short, the receiver makes the noisy observation $\mathbf y = \mathbf H
\mathbf x + \mathbf w$ of a vector of constellation symbols $\mathbf x$
through a channel matrix $\mathbf H$.
Knowing $\mathbf H$ and the variance $N_0$ of the elements of the noise vector $\mathbf w$, the equalizer must produce bitwise \glspl{llr} for a channel decoder.
Equalizers are ranked by the \gls{nve}~\cite{gruber2017deep}, the ratio of the coded
\gls{bler} achieved by the equalizer to that of a reference equalizer, averaged
over a set of \gls{snr} values $\mathcal{S}$,
\begin{equation}
    \mathrm{NVE} = \frac{1}{|\mathcal{S}|}\sum_{\mathrm{SNR}\in\mathcal{S}}
    \frac{\mathrm{BLER}_{\mathrm{cand}}(\mathrm{SNR})}{\mathrm{BLER}_{\mathrm{ref}}(\mathrm{SNR})},
    \label{eq:nve}
\end{equation}
so that $\mathrm{NVE}\le 1$ means the equalizer matches or beats the
reference on this averaged metric. The latency of the equalizer (which must be compatible with \texttt{torch.compile}) is used as a proxy for complexity.

\subsection{System Model}
\label{sec:otfs-system-model}
We consider an \gls{otfs} system whose transmit signal is, for every sub-block,
an \gls{ofdm} waveform with a \gls{cp}. The data symbols $x[l,k]$ are defined on
the \gls{dd} grid, for $0\le l<M$ and $0\le k<N$, where $M$ and $N$ denote the delay and Doppler bin count, respectively. The data symbols are mapped to
the time--frequency grid by the inverse symplectic finite Fourier transform
\begin{equation}
X_{\rm tf}[n,m] = \frac{1}{\sqrt{MN}}\sum_{l=0}^{M-1}\sum_{k=0}^{N-1} x[l,k]\,e^{\,j2\pi\!\left(\frac{nk}{N}-\frac{ml}{M}\right)}
\end{equation}
and then modulated onto a \gls{cp}-\gls{ofdm} waveform with a rectangular
transmit pulse, \gls{cp} length $T_{\rm cp}$, subcarrier spacing $\Df$, symbol period $T=1/\Df$, and sample
period $\Ts{=}T/M$.
We assume a linear time-varying multipath channel with $P$ paths
\begin{equation}
 h(\tau,\nu)=\sum_{i=1}^{P} h_i\,\delta(\tau-\tau_i)\,\delta(\nu-\nu_i),
\end{equation}
where path $i$ has gain $h_i$, delay $\tau_i$, and Doppler $\nu_i$. The
\gls{cp} is assumed longer
than the delay spread and the path parameters $h_i$, $\tau_i$, and $\nu_i$ are constant over the
\gls{otfs} frame.

As shown in Appendix~\ref{sec:otfs-io}, one can derive the following input--output relation for the \gls{otfs} system:
\begin{equation}
  \mathbf y = \mathbf H\,\mathbf x + \mathbf w,
  \label{eq:matrix}
\end{equation}
where $\xv,\yv\in\CC^{MN}$ are the vectorized transmitted and received symbols, respectively, $\mathbf H\in\CC^{MN\times MN}$ is the effective \gls{dd} channel matrix with elements given by \eqref{eq:Hentry}, and $\mathbf w\in\CC^{MN}$ is a vector of complex \gls{awgn} with variance $N_0$.
As discussed at greater length in the appendix, one may truncate the channel matrix to a row-sparse approximation $\widetilde{\mathbf H}$, e.g., by keeping only the \topk\ strongest taps per row, with little loss in accuracy. This structure can be efficiently exploited by several classes of equalizers.

Since Sionna~\cite{sionna} does not support \gls{otfs} in its current version, we provide a new Sionna \gls{otfs} module with all required functionality in the article's code repository.

\subsection{Prior Art}
\label{sec:prior-art}

Most practical \gls{otfs} equalizers are iterative schemes that operate on the
row-sparse $\widetilde{\mathbf H}$ channel representation instead of the dense
$\mathbf H$ matrix. This avoids the prohibitively high complexity of inverting
a large $MN\times MN$ matrix, as in, for example, the \gls{lmmse} estimate $\widehat{\mathbf x}=(\mathbf H\htp\mathbf H+N_0\mathbf I)^{-1}\mathbf H\htp\mathbf y$.
The interference-cancellation \gls{mp}
detector of~\cite{raviteja2018} is the canonical example,
exchanging beliefs over the discrete constellation symbols on the factor graph induced by
the sparse channel. Gaussian-message variants reduce the per-edge cost while
retaining calibrated soft outputs, e.g., the \gls{uamp} detector~\cite{yuan2022uamp} and the \gls{ep} detector~\cite{li2021ep}---which serves as the reference baseline in this
work due to its excellent performance and low complexity. All of these exploit the row sparsity of
$\widetilde{\mathbf H}$, so their per-iteration complexity scales as $MNS$, where $S$ is
the number of retained taps per row, rather than as $(MN)^2$.
Low-complexity \gls{lmmse} receivers in~\cite{tiwari2019,surabhi2020} exploit a
\emph{full} block-circulant-with-circulant-blocks effective \gls{dd} channel
matrix, which reduces equalization to a 2-D \gls{dft}. That stronger property
holds for ideal biorthogonal pulses with integer or slowly varying Doppler, but
fails under the rectangular-\gls{cp} waveform with \emph{continuous} delay
\emph{and} Doppler considered here. The channel nevertheless remains
block-circulant along the Doppler index~\cite{raviteja2018}, so that a Doppler \gls{dft}
still decouples the $MN\times MN$ system into $N$ independent $M\times M$
blocks. Most competitive algorithms discovered in this task exploit this structure.

\subsection{Simulation Setup}
We consider an \gls{otfs} system with $M=N=64$ delay and Doppler bins, a carrier
frequency of \qty{4}{\giga\hertz}, a subcarrier spacing of $\Df=\qty{15}{\kilo\hertz}$, and a $16$-sample \gls{cp}.
Information bits are encoded by a rate-$1/2$ 5G \gls{ldpc} code with block
length \num{16384}, mapped to a
$16$-\gls{qam} constellation, and then placed on the \gls{dd} grid. As perfect
\gls{csi} is assumed, no pilots need to be sent.

We use the continuous \gls{dd} channel model \eqref{eq:io} (see Appendix~\ref{sec:otfs-io}) with
$P=6$ paths. Per frame, the path gains are independently and identically distributed according to $\CN(0,1/P)$, the delays are
uniform on $[0,\,14\,\Ts]$ (i.e., up to \qty{14.6}{\micro\second}), and the Doppler shifts are uniform on
$[-0.5\Df,\,0.5\Df]$ (i.e., from \qty{-7.5}{\kilo\hertz} to \qty{7.5}{\kilo\hertz}). The first path is fixed at zero delay and zero Doppler. The path parameters are
constant over the frame and the exact full-period input--output relation is
used for the simulation of $\mathbf y$. An equalizer can use either the
dense representation or a row-sparse version obtained by retaining the \topk\ strongest taps per row, where the number of retained taps is a free parameter.
The noise variance follows the unit-average-symbol-energy
convention $N_0=10^{-\mathrm{SNR}_{\mathrm{dB}}/10}$, and the evaluation
\gls{snr} grid for the \gls{nve} calculation is $\mathcal{S}=\{\qty{13}{\decibel},\qty{16}{\decibel}\}$, with \gls{ep}~\cite{li2021ep} ($\topk = 256$) as the reference equalizer.

\subsection{Results}
\label{sec:otfs-results}

\rowcolors{2}{gray!12}{white}
\begin{table*}[t]
  \centering
  \caption{Algorithm families explored in the candidate search for the \gls{otfs} equalizer design task}
  \label{tab:families}
  \renewcommand{\arraystretch}{1.2}
  \setlength{\tabcolsep}{4pt}
  \begin{tabularx}{\textwidth}{l X c c c} 
    \toprule
    \rowcolor{white}\textbf{Family} & \textbf{Core operation} & \textbf{Candidates} &
    \textbf{Best NVE} & \textbf{Latency of best} \\
    \midrule
    \acrshort{vamp}\,/\,\acrshort{oamp}~\cite{schniter2016vamp,ma2017oamp} &
      Block-FFT linear module with extrinsic Gaussian--QAM iterations &
      \num{557} & \num{0.496} & ~\qty{5.28}{\milli\second} \\
    \acrshort{ep}\,/\,\acrshort{ec}~\cite{li2021ep,opper2005ec} &
      Linear Gaussian updates alternating with extrinsic QAM denoiser &
      \num{392} & \num{0.496} & ~\qty{6.40}{\milli\second} \\
    Proximal \acrshort{map}\,/\,\acrshort{admm}~\cite{combettes2011proximal,boyd2011admm} &
      Projected-gradient or ADMM steps with QAM proximal thresholding &
      \num{116} & \num{0.549} & ~\qty{6.56}{\milli\second} \\
    List\,/\,$K$-best~\cite{hassibi2005sphere,wong2002kbest} &
      Local search over a small QAM neighborhood with residual scoring &
      ~\num{88} & \num{0.561} & ~\qty{3.36}{\milli\second} \\
    \acrshort{gamp}\,/\,\acrshort{amp}~\cite{rangan2011gamp,donoho2009amp} &
      Scalar AMP on the sparse DD graph with Onsager correction &
      \num{108} & \num{0.630} & \qty{37.60}{\milli\second} \\
    Linear\,/\,\acrshort{lmmse}~\cite{tiwari2019,surabhi2020} &
      One-shot Wiener solve, dense or Doppler-FFT block-structured &
      \num{252} & \num{0.659} & ~\qty{2.24}{\milli\second} \\
    \acrshort{bp}\,/\,\acrshort{mp}~\cite{raviteja2018} &
      Sparse-graph belief propagation with leave-one-out interference &
      ~\num{97} & \num{0.677} & \qty{19.52}{\milli\second} \\
    \acrshort{pic}\,/\,\acrshort{sic}~\cite{raviteja2018} &
      Parallel or successive interference cancellation on sparse residuals &
      \num{115} & \num{0.677} & \qty{10.64}{\milli\second} \\
    Krylov\,/\,\acrshort{pcg}~\cite{saad2003iterative} &
      Iterative solve of the regularized normal equations &
      \num{130} & \num{1.142} & \qty{10.64}{\milli\second} \\
    Fusion\,/\,gating~\cite{jacobs1991moe} &
      Weighted combination of multiple detector LLR streams &
      ~\num{22} & \num{1.258} & ~\qty{3.28}{\milli\second} \\
    Low-rank\,/\,subspace~\cite{golub2013matrix} &
      Reduced-dimension solve via low-rank channel approximation &
      ~\num{13} & \num{1.823} & ~\qty{5.60}{\milli\second} \\
    \bottomrule
  \end{tabularx}
  \par\vspace{6pt}
  \parbox{\textwidth}{\fontsize{8pt}{9.5pt}\linespread{1.0}\selectfont
  \emph{Abbreviations:}
    \acrfull{vamp}, \acrfull{oamp},
    \acrfull{ep}, \acrfull{ec},
    \acrfull{map}, \acrfull{admm},
    \acrfull{gamp}, \acrfull{amp},
    \acrfull{lmmse},
    \acrfull{bp}, \acrfull{mp},
    \acrfull{pic}, \acrfull{sic},
    \acrfull{pcg}.}
\end{table*}
\rowcolors{0}{}{}

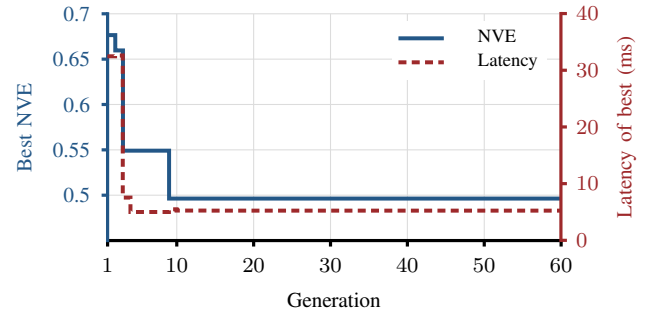
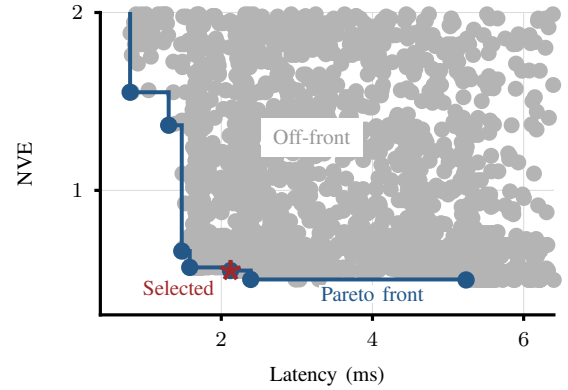
\begin{figure}
  \centering
  \subfloat[Best \gls{nve} across generations and their corresponding latency]{\begin{tikzpicture}
    \begin{axis}[standard_plotstyle,
	        width=6cm, height=3cm,
        xmin=1, xmax=60, ymin=0.45, ymax=0.7,
		every outer y axis line/.append style={black, very thick, cap=rect},
		every outer x axis line/.append style={black, very thick, cap=rect},   
		axis y line*=left,
        xlabel={Generation},
        ylabel={Best NVE},
        ylabel style={myblue},
        yticklabel style={myblue, text width=0.5cm, align=right},
        xtick={1,10,20,30,40,50,60},
        ytick={0.5,0.55,0.6,0.65,0.7},
        yticklabels={0.5,0.55,0.6,0.65,0.7},
        yminorgrids=false,
        grid style={mylightgray4},
        every outer y axis line/.append style={myblue},
        y tick style={myblue, very thick},     
    ]
        \addplot[myblue, const plot, mark=none, line width=1.5pt]
            table[col sep=comma, x expr=\thisrow{generation}+1, y=best_metric]{data/otfs_leaderboard_evolution.csv};
    \end{axis}
    \begin{axis}[
        width=6cm, height=3cm, scale only axis,
		every outer y axis line/.append style={black, very thick, cap=rect},
		every outer x axis line/.append style={black, very thick, cap=rect},   
		xmin=1, xmax=60, ymin=0, ymax=40,
        axis x line=none,
        axis y line*=right,
        ytick pos=right,
        scaled y ticks=false,
        ylabel={Latency of best (\unit{\milli\second})},
        ylabel style={myred, font=\small},
        yticklabel style={myred, font=\footnotesize},
        ytick={0,10,20,30,40},
        every outer y axis line/.append style={myred, very thick},
        tick align=outside,
        y tick style={myred, very thick},
        major tick length=2.25pt,
        legend pos=north east,
        legend columns=1,
        legend style={font=\scriptsize, column sep=1em,
                      draw=none, fill=white, fill opacity=0.75, text opacity=1,
                      row sep=-1pt, inner sep=1.5pt},
        legend cell align=left,
    ]
        \addplot[myred, const plot, line width=1.5pt, densely dashed, mark=none, forget plot]
            table[col sep=comma, x expr=\thisrow{generation}+1, y expr=8*\thisrow{best_complexity}]{data/otfs_leaderboard_evolution.csv};
        \addlegendimage{myblue, line width=1.5pt}
        \addlegendentry{NVE}
        \addlegendimage{myred, densely dashed, line width=1.5pt}
        \addlegendentry{Latency}
    \end{axis}
\end{tikzpicture}\label{fig:otfs_leaderboard_evolution}}\\[4pt]
  \subfloat[Generated equalizers in the latency--\gls{nve} plane]{\begin{tikzpicture}
    \begin{axis}[standard_plotstyle,
        width=6cm, height=4cm,
        xlabel={Latency (\unit{\milli\second})},
        ylabel={NVE},
        ylabel style={black, font=\small},
        xlabel style={black, font=\small},
        yticklabel style={black, font=\footnotesize, text width=0.4cm, align=right},
        scaled x ticks=false,
        xmin=0.4, xmax=6.4,
        ymin=0.3, ymax=2,
        ytick={1,2,3,4},
        restrict x to domain=0:8,
        restrict y to domain=0:3.5,
        unbounded coords=jump,
		every outer y axis line/.append style={black, very thick, cap=rect},
		every outer x axis line/.append style={black, very thick, cap=rect},        
        grid style={mylightgray4},
        legend style={
            font=\scriptsize, draw=mylightgray3, fill=white,
            at={(0.97,0.97)}, anchor=north east, row sep=-1pt,
        },
        legend cell align=left,
    ]
		\addplot[only marks, mark=*, mark size=2pt, color=mylightgray3, opacity=0., mark options={fill opacity=0.7}]
    	table[col sep=comma, x expr=8*\thisrow{complexity}, y=metric]{data/otfs_leaderboard_off_front.csv};

        \addplot[only marks, myblue, mark size=2.25pt, opacity=0., mark options={fill opacity=1}]
            table[col sep=comma, x expr=8*\thisrow{complexity}, y=metric]{data/otfs_leaderboard_pareto_front.csv};

        \addplot[myblue, const plot mark right, line width=1.5pt]
            table[col sep=comma, x expr=8*\thisrow{complexity}, y=metric]{data/otfs_leaderboard_pareto_front.csv};

        \addplot[myred, only marks, mark=star, mark size=4pt, ultra thick]
            coordinates {(2.1224, 0.5491)};
            
         \draw[white] (7.3,1.0) -- (7.3,1.5);
         
  		\node[mylightgray2, labelstyle, inner sep=1.6mm] (d) at (3.2, 1.3) {Off-front};
  		\node[myblue, labelstyle] (d) at (4.0, 0.42) {Pareto front};
  		\node[myred, labelstyle] (d) at (1.44, 0.45) {Selected};
  		         
    \end{axis}
\end{tikzpicture}\label{fig:otfs_leaderboard_pareto}}
  \caption{Results of the \gls{aite} run for the \gls{otfs} equalizer task.}
  \label{fig:otfs_leaderboard}
\end{figure}

Over \num{60} generations, \gls{aite} produced \num{1890} working candidates
implementing \num{1114} distinct equalizer architectures. We used GPT-5.5 as the underlying \gls{llm} and a cluster of \num{16} \glspl{gpu} running \num{32} workers per generation. The search broadly explored the accuracy--latency tradeoff across a large
variety of algorithm families cataloged in Table~\ref{tab:families}.
Fig.~\ref{fig:otfs_leaderboard}\subref{fig:otfs_leaderboard_evolution} shows the evolution of the best
\gls{nve} across generations and the corresponding latency.
The best \gls{nve} decreases from $0.677$ to $0.549$ by generation~3 and
reaches $0.496$ by generation~9, with no further improvement observed
thereafter. The search then focused on ways to reduce latency without degrading performance.
Fig.~\ref{fig:otfs_leaderboard}\subref{fig:otfs_leaderboard_pareto} shows the Pareto front together with
off-front solutions, zoomed to the low-\gls{nve}-low-latency region for readability.
Notably, many families of algorithms reach comparable performance,
especially with the right hyperparameter tuning. All well-performing
algorithms exploit the block-circulant channel structure mentioned earlier.

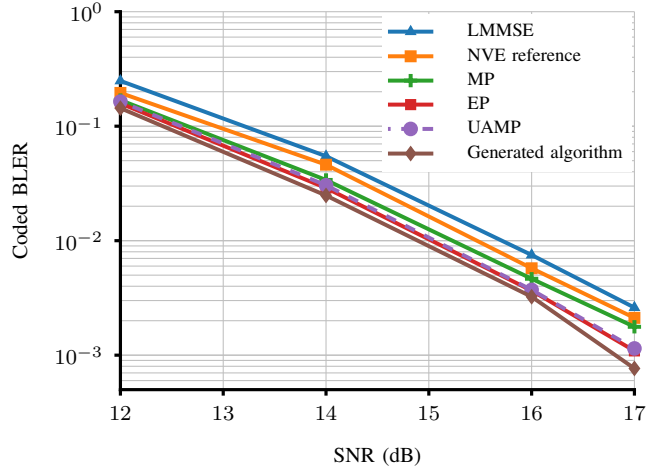
\begin{figure}
  \centering
  \begin{minipage}{\columnwidth}
    \centering
    \begin{tikzpicture}
\begin{semilogyaxis}[ber_plotstyle,
width=6.8cm,
height=5cm,
xmin=12,
xmax=17,
ymin=0.0005,
ymax=1,
ylabel=Coded BLER,
line width=1.5pt,
mark size=2pt,
mark options={solid, line width=0.6pt},
cycle list name=matplotlib,
legend columns=1,
legend style={at={(0.999,0.999)}, anchor=north east,
              font=\scriptsize, column sep=1em,
              draw=none, fill=white, fill opacity=0.75, text opacity=1,
              row sep=-1pt, inner sep=1.5pt},
legend cell align=left,
]

\addplot+[mark=triangle*,]
    table[col sep=comma, x=snr, y=lmmse]{data/otfs_bler.csv};
\addlegendentry{LMMSE}

\addplot+[mark=square*, mark options={solid}, mark size=1.5pt]
    table[col sep=comma, x=snr, y=ep256]{data/otfs_bler.csv};
\addlegendentry{NVE reference}

\addplot+[mark=+, mark size=2.5pt, mark options={line width=1.5pt}]
    table[col sep=comma, x=snr, y=mp]{data/otfs_bler.csv};
\addlegendentry{MP}

\addplot+[mark=square*, mark size=1.8pt]
    table[col sep=comma, x=snr, y=ep]{data/otfs_bler.csv};
\addlegendentry{EP}

\addplot+[dash pattern={on 4pt off 4pt}, dash phase=3pt, mark=*, mark options={fill opacity=0, solid}]
    table[col sep=comma, x=snr, y=uamp]{data/otfs_bler.csv};
\addlegendentry{UAMP}

\addplot+[mark=diamond*, mark size=2.5pt]
    table[col sep=comma, x=snr, y=agentic]{data/otfs_bler.csv};
\addlegendentry{Generated algorithm}

\end{semilogyaxis}
\end{tikzpicture}
    \vspace{-3mm}
  \end{minipage}
  \caption{Coded \gls{bler} versus \gls{snr} for various \gls{otfs} equalizers.
  The values in parentheses show the speedup relative to \gls{lmmse} on an NVIDIA~RTX~PRO~6000.
  \gls{lmmse} uses the dense channel representation; \gls{mp} uses $\topk = 512$; \gls{uamp} uses $\topk = 2048$;
  \gls{ep} uses $\topk = 512$; and the generated algorithm uses $\topk = 2560$.
  The equalizer used as a reference for the \gls{nve} in \eqref{eq:nve} is \gls{ep} with $\topk = 256$.
  }
  \label{fig:otfs_bler}
\end{figure}

We select a candidate from the \gls{ec} family found in the 59th generation with an interesting
accuracy--latency tradeoff ($\mathrm{NVE}=0.549$,
latency \qty{2.17}{\milli\second}), highlighted in
Fig.~\ref{fig:otfs_leaderboard}\subref{fig:otfs_leaderboard_pareto}, as a representative example for
comparison with several baselines. 
As shown in Fig.~\ref{fig:otfs_bler}, the generated algorithm
beats all conventional baselines in coded \gls{bler}, including \gls{mp}~\cite{raviteja2018},
\gls{ep}~\cite{li2021ep}, and \gls{uamp}~\cite{yuan2022uamp}. At the same time, Table~\ref{tab:otfs_latency} shows a
per-frame latency of \qty{2.17}{\milli\second} on an NVIDIA~RTX~PRO~6000 and \qty{3.52}{\milli\second} on an
NVIDIA~DGX~Spark. These values correspond to speedups of around $3.6\times$ over the most competitive baseline, \gls{uamp}~\cite{yuan2022uamp}, on both platforms and of $66\times$ and $210\times$ over \gls{lmmse} on the RTX~PRO~6000 and DGX~Spark, respectively. Note that all baselines underwent the same Optuna-based hyperparameter tuning~\cite{optuna} as the generated algorithms, and their implementations were optimized for latency with the help of agentic coding tools.
A detailed discussion of the generated algorithm and how it achieves state-of-the-art
performance at lower latency is deferred to
Appendix~\ref{sec:otfs-candidate-description}. This fine-grained optimization was performed across hundreds of candidates explored in parallel, illustrating how evolutionary agentic search can apply implementation-level refinement at scale.

\begin{table}
  \centering
  \renewcommand{\arraystretch}{1.1}
  \caption{Per-frame inference latency. Speedup
  is relative to \gls{lmmse}}
  \label{tab:otfs_latency}
  \footnotesize
  \newcommand{\LatRTX}{143.84}
  \newcommand{\LatDGX}{741.92}
  \begin{tabular*}{\columnwidth}{@{\hspace{2pt}} l @{\extracolsep{\fill}} c @{\extracolsep{0pt}\hspace{2\tabcolsep}} c @{\extracolsep{\fill}} c @{\extracolsep{0pt}\hspace{2\tabcolsep}} c @{\hspace{2pt}}}
	  \toprule
      & \multicolumn{2}{c}{RTX~PRO~6000} & \multicolumn{2}{c}{DGX~Spark} \\
      \textbf{Equalizer}~ & \textbf{Latency} & \textbf{Speedup}~ & \textbf{Latency} & \textbf{Speedup} \\
      \midrule
      \gls{lmmse}  & \qty{\LatRTX}{\milli\second} & ~\speedup{\LatRTX}{\LatRTX} & \qty{\LatDGX}{\milli\second} & ~~\speedup{\LatDGX}{\LatDGX} \\
      \gls{mp}     & ~\qty{58.56}{\milli\second}  & ~\speedup{\LatRTX}{58.56}  & \qty{365.92}{\milli\second} & ~~\speedup{\LatDGX}{365.92} \\
      \gls{ep}     & ~\qty{18.64}{\milli\second}  & ~\speedup{\LatRTX}{18.64}  & \qty{108.00}{\milli\second} & ~~\speedup{\LatDGX}{108} \\
      \gls{uamp}   & ~~\qty{7.76}{\milli\second}   & \speedup{\LatRTX}{7.76}   & ~\qty{12.16}{\milli\second}  & ~\speedup{\LatDGX}{12.16}  \\
      Generated    & ~~\qty{2.17}{\milli\second}   & \speedup{\LatRTX}{2.17}   & ~~\qty{3.52}{\milli\second}   & \speedup{\LatDGX}{3.52}   \\
      \bottomrule
  \end{tabular*}
\end{table}

\section{Explainable Pilotless OFDM Receiver}
\label{sec:pilotless}

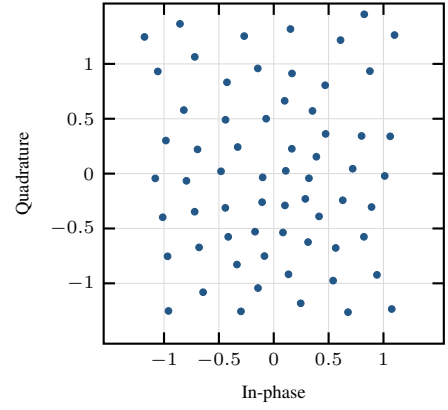
\begin{figure}[t]
    \centering
    \vspace{3mm}
    \begin{tikzpicture}
    \begin{axis}[
        width=4.5cm, height=4.5cm,
        scale only axis,
        axis equal image,
        xmin=-1.55, xmax=1.55,
        ymin=-1.55, ymax=1.55,
        axis background/.style={fill=none},
        axis lines=box,
        axis line style={black, thick, -},
        xlabel={In-phase}, ylabel={Quadrature},
        label style={font=\scriptsize, black},
        xtick={-1,-0.5,0,0.5,1}, ytick={-1,-0.5,0,0.5,1},
        xmajorgrids, ymajorgrids,
        grid style={mylightgray4},
        tick style={black, thick},
        ticklabel style={font=\scriptsize, black},
        clip mode=individual,
    ]
        \addplot[
            only marks, mark=*, mark size=1.25pt,
            color=myblue,
        ] table {
        x y label
        -0.0693 +0.5003 000000
        +0.4129 -0.3899 000001
        +0.1021 -0.2899 000010
        +0.2878 -0.2304 000011
        +0.0996 +0.6635 000100
        -0.2693 +1.2533 000101
        +0.1663 +0.9128 000110
        -0.1454 +0.9587 000111
        +0.7195 +0.0444 001000
        +0.6290 -0.2427 001001
        +1.0117 -0.0213 001010
        +0.8914 -0.3048 001011
        +0.8765 +0.9341 001100
        +0.6087 +1.2167 001101
        +1.1009 +1.2627 001110
        +0.8250 +1.4518 001111
        +0.0841 -0.5369 010000
        +0.3125 -0.6242 010001
        -0.1061 -0.2603 010010
        -0.1011 -0.0343 010011
        -0.1705 -0.5295 010100
        +0.2458 -1.1820 010101
        -0.0848 -0.7515 010110
        +0.1342 -0.9177 010111
        +0.8004 +0.3433 011000
        +0.5648 -0.6774 011001
        +1.0611 +0.3399 011010
        +0.8218 -0.5761 011011
        +0.9406 -0.9228 011100
        +0.5412 -0.9753 011101
        +1.0754 -1.2341 011110
        +0.6767 -1.2627 011111
        +0.4729 +0.3619 100000
        -0.4395 +0.4906 100001
        +0.3883 +0.1537 100010
        +0.3207 -0.0422 100011
        +0.3531 +0.5713 100100
        +0.1525 +1.3174 100101
        +0.4688 +0.8052 100110
        -0.4263 +0.8322 100111
        -0.7958 -0.0658 101000
        -0.6946 +0.2206 101001
        -1.0796 -0.0436 101010
        -0.9830 +0.3013 101011
        -1.0567 +0.9313 101100
        -0.7195 +1.0635 101101
        -1.1781 +1.2458 101110
        -0.8548 +1.3645 101111
        -0.4412 -0.3120 110000
        -0.3284 +0.2421 110001
        +0.1648 +0.2264 110010
        +0.1098 +0.0254 110011
        -0.4151 -0.5763 110100
        -0.2989 -1.2555 110101
        -0.3354 -0.8287 110110
        -0.1436 -1.0429 110111
        -0.7204 -0.3473 111000
        -0.4806 +0.0211 111001
        -1.0119 -0.3980 111010
        -0.8195 +0.5792 111011
        -0.6804 -0.6723 111100
        -0.6439 -1.0810 111101
        -0.9677 -0.7537 111110
        -0.9594 -1.2516 111111
        };
    \end{axis}
\end{tikzpicture}
    \caption{Learned \(2^M\)-point constellation (\(M=6\)).}
    \label{fig:constellation}
\end{figure}

The second task assigned to \gls{aite} is to generate an explainable pilotless \gls{ofdm} receiver.
This line of work builds on prior results on pilotless \gls{ofdm} communication~\cite{9508784}, which showed that the \gls{ofdm} waveform can operate without reference signals when the transmitter and receiver are jointly optimized through end-to-end learning.
In that setting, a learned constellation shown in Fig.~\ref{fig:constellation} is used to modulate the transmitted bits on all \glspl{re} of the \gls{ofdm} resource grid, and a neural-network-based receiver maps the received grid directly to \glspl{llr} for the coded bits.

The objective of this experiment is to retain the pilotless transmission scheme using the learned constellation, while replacing the black-box neural receiver with an explainable receiver algorithm discovered by \gls{aite}.
Algorithms using sparse pilots exist~\cite{6847742, 6302202, 5585794}, as do fully pilotless schemes relying on multiple constellations to enable channel estimation~\cite{8417804, 8012416}.
However, the literature still lacks a non-neural pilotless \gls{ofdm} method that uses a single constellation over all \glspl{re}.
Moreover, none of the existing solutions seems to address the computation of \glspl{llr} for the transmitted bits, which are required for channel decoding and are challenging to obtain as they rely on accurate calculation of the channel estimation error.
It is thus unlikely that \gls{aite} can reproduce an algorithm from its \gls{llm} training data.

\subsection{System Model}
\label{subsec:pilotless_system}

We consider a single-input single-output \gls{ofdm} link operating on a resource grid with \(N_F\) subcarriers and \(N_T\) \gls{ofdm} symbols per slot, yielding \(N_F N_T\) \glspl{re} indexed by \((f,t)\) with \(1 \leq f \leq N_F\) and \(1 \leq t \leq N_T\).
The constellation used to modulate the transmitted bits is denoted by $\Cc \subset \CC$, with $\abs{\Cc} = 2^M$, where $M$ is the number of bits per symbol.

\begin{figure}[t]
    \centering
    \subfloat[With \gls{dmrs}]{
\begin{tikzpicture}[
    pilot/.style={thick, draw=mygray, fill=mygray, fill opacity=0.25, minimum width=0.22cm, minimum height=0.18cm, inner sep=0pt},
    data/.style={thick, draw=myaltgray, fill=myaltgray, fill opacity=0.25, minimum width=0.22cm, minimum height=0.18cm, inner sep=0pt},
    axlbl/.style={font=\footnotesize},
]
    \def\dx{0.23}
    \def\dy{0.185}
    \foreach \t in {0,...,13} {
        \foreach \f in {0,...,11} {
            \ifnum\t=2
            \else\ifnum\t=11
            \else
                \node[data] at (\f*\dx, -\t*\dy) {};
            \fi\fi
        }
    }
    \foreach \t in {0,...,13} {
        \foreach \f in {0,...,11} {
            \ifnum\t=2
                \node[pilot] at (\f*\dx, -\t*\dy) {};
            \else\ifnum\t=11
                \node[pilot] at (\f*\dx, -\t*\dy) {};
            \fi\fi
        }
    }    
    \node[axlbl, rotate=90] at (-0.5, -6.5*\dy) {OFDM symbols};
    \node[axlbl] at (5.5*\dx, -14*\dy - 0.2) {Subcarriers};
\end{tikzpicture}}%
    \hspace{1cm}
    \subfloat[Pilotless (all data)]{
\begin{tikzpicture}[
    data/.style={thick, draw=myaltgray, fill=myaltgray, fill opacity=0.25, minimum width=0.22cm, minimum height=0.18cm, inner sep=0pt},
    axlbl/.style={font=\footnotesize},
]
    \def\dx{0.23}
    \def\dy{0.185}
    \foreach \t in {0,...,13} {
        \foreach \f in {0,...,11} {
            \node[data] at (\f*\dx, -\t*\dy) {};
        }
    }
    \node[axlbl, rotate=90] at (-0.5, -6.5*\dy) {OFDM symbols};
    \node[axlbl] at (5.5*\dx, -14*\dy - 0.2) {Subcarriers};
\end{tikzpicture}\label{fig:rg_grid_nopilots}}
    \caption{Resource grids (\(12 \times 14\) \glspl{re} in this figure) for pilot-assisted and pilotless transmission. Dark \glspl{re} carry \gls{dmrs}; light \glspl{re} carry data. In~(a), \gls{dmrs} symbols are placed in \gls{ofdm} symbols \(t \in \{3,12\}\).}
    \label{fig:rg_comparison}
\end{figure}
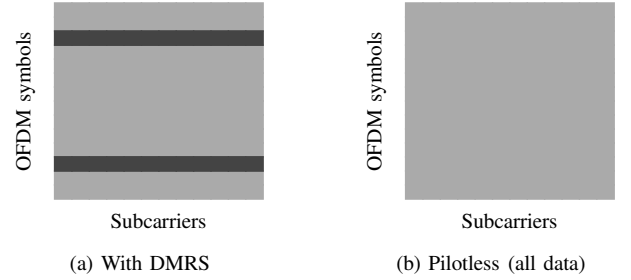

Fig.~\ref{fig:rg_comparison} contrasts the resource allocation in a conventional pilot-assisted grid and in the pilotless grid.
In the former case, a subset \(\Pc\) of \glspl{re} carries known \gls{dmrs} symbols.
In the latter case, \(\Pc = \emptyset\) and all \glspl{re} are occupied by data symbols.

On the receiver side, after cyclic-prefix removal and \gls{dft}, the complex baseband received symbols satisfy
\begin{equation}
    \Ym = \Xm \odot \Hm + \Wm,
    \label{eq:ofdm_rx_model}
\end{equation}
where \(\Xm \in \CC^{N_F \times N_T}\) are the transmitted symbols, \(\Hm \in \CC^{N_F \times N_T}\) is the channel matrix, \(\Wm \in \CC^{N_F \times N_T}\) is \gls{awgn} with per-\gls{re} variance \(N_0\), \(\Ym \in \CC^{N_F \times N_T}\) is the received grid, and \(\odot\) denotes elementwise multiplication.

\begin{figure}[t]
    \centering
    \input{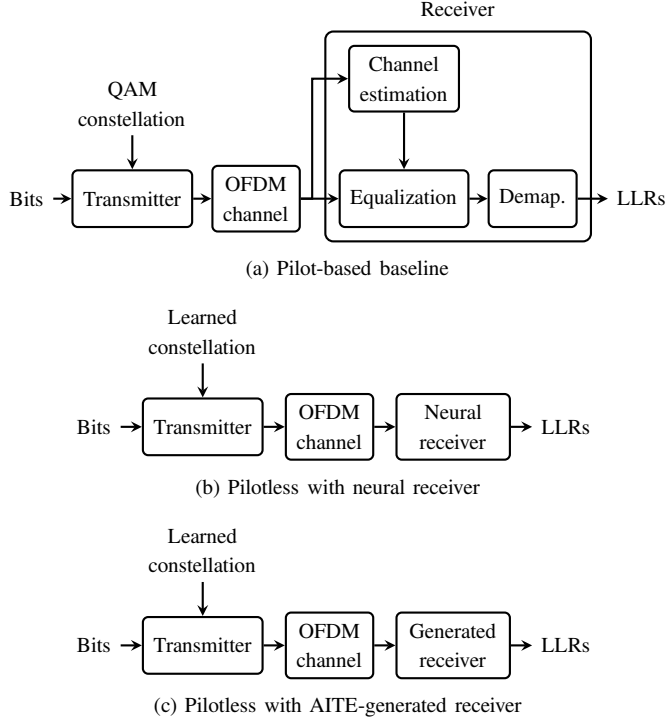}
    \caption{End-to-end \gls{ofdm} receivers considered in this section: (a)~conventional pilot-based baseline with \gls{qam} and \gls{dmrs}; (b)~pilotless system with learned constellation and neural receiver~\cite{9508784}; (c)~pilotless system with learned constellation and \gls{aite}-generated receiver.}
    \label{fig:pilotless_systems}
\end{figure}

The pilotless system we build upon was introduced in~\cite{9508784}, in which the constellation at the transmitter is jointly optimized with a neural network at the receiver, as shown in Fig.~\ref{fig:pilotless_systems}\subref{fig:pilotless_sys_b}.
The neural receiver substitutes for channel estimation, equalization, and demapping. It maps the received resource grid to \glspl{llr}. In this work, the neural receiver is a residual convolutional neural network as in~\cite{9508784}.
On the transmitter side, only the constellation is learned, while the remainder of the transmission pipeline stays conventional.
The trainable parameters consist of a set of \(2^M\) complex numbers \(\widetilde{\Cc} = \{\widetilde{c}_1, \dots, \widetilde{c}_{2^M}\}\), which are centered and normalized to form the constellation \(\Cc\) used to modulate the transmitted data,
\begin{equation}
    \Cc = \LP \frac{\widetilde{c} - \mu}{\sqrt{\frac{1}{2^M}\sum_{\widetilde{c}' \in \widetilde{\Cc}} \abs{\widetilde{c}'}^2 - \abs{\mu}^2}} \quad \Bigg\lvert \quad \widetilde{c} \in \widetilde{\Cc}\RP,
    \label{eq:learned_constellation}
\end{equation}
where \(\mu = \frac{1}{2^M}\sum_{\widetilde{c} \in \widetilde{\Cc}} \widetilde{c}\), so that \(\Cc\) has zero mean and unit average power.
Centering prevents the learned constellation from embedding a superimposed pilot in its mean, which would act as a known reference signal.
The learned constellation is applied to every \gls{re} of the resource grid, as shown in Fig.~\ref{fig:rg_comparison}\subref{fig:rg_grid_nopilots}. Consequently, the resulting scheme uses neither orthogonal nor superimposed pilots.

The results reported in~\cite{9508784} show that the neural receiver is able to recover the transmitted bits despite the lack of pilots.
Removing pilots frees the \glspl{re} that conventional systems reserve for reference signals and reallocates them to data, which translates into significant throughput gains.
Achieving this requires the learned constellation to adopt an unconventional geometry, shown in Fig.~\ref{fig:constellation}.
While effective, the neural receiver inherits the general lack of explainability of neural networks. This makes the design difficult to analyze and extend, e.g., to multiple spatial streams or different modulation orders. It also motivates the search for an explainable algorithmic receiver that \gls{aite} is asked to find.

\subsection{Simulation Setup}
\label{subsec:simulation_setup}

We consider an \gls{ofdm} system with \(N_F = 72\) subcarriers and \(N_T = 14\) \gls{ofdm} symbols.
To train the end-to-end learning system (Fig.~\ref{fig:pilotless_systems}\subref{fig:pilotless_sys_b}) and generate algorithms with \gls{aite} (Fig.~\ref{fig:pilotless_systems}\subref{fig:pilotless_sys_c}), we use the TDL-C channel model with a nominal delay spread of \qty{100}{\nano\second}, a carrier frequency of \qty{2.6}{\giga\hertz}, a subcarrier spacing of \qty{30}{\kilo\hertz}, and user speeds between 0 and \qty{3}{\meter\per\second}.
The constellation optimized through end-to-end learning, shown in Fig.~\ref{fig:constellation}, is reused unchanged for generating algorithms with \gls{aite}.
The evaluation tool returns the \gls{nve} defined in~\eqref{eq:nve}, computed over two \gls{snr} points, \qty{15}{\decibel} and \qty{20}{\decibel}, using the pilot-based baseline in Fig.~\ref{fig:pilotless_systems}\subref{fig:pilotless_sys_a} as the reference.
The average latency required to run the algorithm is used as a proxy for complexity.

We benchmark the generated algorithm on the TDL-A and TDL-D models, which are \gls{nlos} and \gls{los} models, respectively, with the same nominal delay spread and user speeds.
Evaluating on channel models unseen during training probes the generalization capabilities of the receivers.
For all receivers considered in this section, the resulting \gls{llr} vector is passed to the same \gls{ldpc} decoder with code rate \(r_c = 0.7\), which outputs hard decisions on the transmitted information bits.

\subsection{Results}
\label{subsubsec:pilotless_results}

\begin{figure}[t]
    \centering
    \subfloat[Best \gls{nve} achieved so far and their corresponding latency]{\begin{tikzpicture}
    \begin{axis}[standard_plotstyle,
	        width=6cm, height=3cm,
        xmin=1, xmax=42, ymin=0.5, ymax=5,
		every outer y axis line/.append style={black, very thick, cap=rect},
		every outer x axis line/.append style={black, very thick, cap=rect},        
		axis y line*=left,
        xlabel={Generation},
        ylabel={Best NVE},
        ylabel style={myblue},
        yticklabel style={myblue, text width=0.3cm, align=right},
        xtick={1,10,20,30,40},
        ytick={1,2,3,4,5},
        yminorgrids=false,
        grid style={mylightgray4},
        every outer y axis line/.append style={myblue},
        y tick style={myblue},
    ]
        \addplot[myblue, const plot, mark=none, line width=1.5pt]
            table[col sep=comma, x expr=\thisrow{generation}+1, y=best_metric]{data/pilotless_leaderboard_evolution.csv};
    \end{axis}
    \begin{axis}[
        width=6cm, height=3cm, scale only axis,
        xmin=1, xmax=42, ymin=0, ymax=2.5,
		every outer y axis line/.append style={black, very thick, cap=rect},
		every outer x axis line/.append style={black, very thick, cap=rect},  
		axis x line=none,
        axis y line*=right,
        ytick pos=right,
        scaled y ticks=false,
        ylabel={Latency of best (\unit{\milli\second})},
        ylabel style={myred, font=\small},
        yticklabel style={myred,  font=\footnotesize},
        ytick={0, 0.5, 1, 1.5, 2, 2.5},
        yticklabels={0, 0.5, 1, 1.5, 2, 2.5},
        every outer y axis line/.append style={myred, very thick},
        tick align=outside,
        y tick style={myred, very thick},
        major tick length=2.25pt,
        legend columns=1,
        legend style={font=\scriptsize, column sep=1em,
                      draw=none, fill=white, fill opacity=0.75, text opacity=1,
                      row sep=-1pt, inner sep=1.5pt,
                      at={(0.99,1)}, anchor=north east},
        legend cell align=left,
    ]
        \addplot[myred, const plot, line width=1.5pt, densely dashed, mark=none, forget plot]
            table[col sep=comma, x expr=\thisrow{generation}+1, y expr=\thisrow{best_complexity}*1000]{data/pilotless_leaderboard_evolution.csv};
        \addlegendimage{myblue, line width=1.5pt}
        \addlegendentry{NVE}
        \addlegendimage{myred, densely dashed, line width=1.5pt}
        \addlegendentry{Latency}
    \end{axis}
\end{tikzpicture}\label{fig:leaderboard_evolution}}\\[4pt]
    \subfloat[Generated algorithms in the latency--\gls{nve} plane]{\hspace{-1cm}\begin{tikzpicture}
    \begin{axis}[standard_plotstyle,
        width=6cm, height=4cm,
        xlabel={Latency (\unit{\milli\second})},
        ylabel={NVE},
        ylabel style={black, font=\small},
        xlabel style={black, font=\small},
        yticklabel style={black, font=\footnotesize, text width=0.4cm, align=right},
        scaled x ticks=false,
        xmin=0.2, xmax=2,
        ymin=0.7, ymax=4,
        ytick={1,2,3,4},
        restrict y to domain=0:4.5,
        unbounded coords=jump,
        grid style={mylightgray4},
		every outer y axis line/.append style={black, very thick, cap=rect},
		every outer x axis line/.append style={black, very thick, cap=rect},    
		legend style={
            font=\scriptsize, draw=mylightgray3, fill=white,
            at={(0.97,0.97)}, anchor=north east, row sep=-1pt,
        },
        legend cell align=left,
    ]

		\addplot[only marks, mark=*, mark size=2pt, color=mylightgray3, opacity=0., mark options={fill opacity=0.7}]
            table[col sep=comma, x expr=\thisrow{complexity}*1000, y=metric]{data/pilotless_leaderboard_off_front.csv};

        \addplot[only marks, myblue, mark size=2.25pt, opacity=0., mark options={fill opacity=1}]
            table[col sep=comma, x expr=\thisrow{complexity}*1000, y=metric]{data/pilotless_leaderboard_pareto_front.csv};
            
        \addplot[myblue, const plot mark right, line width=1.5pt]
            table[col sep=comma, x expr=\thisrow{complexity}*1000, y=metric]{data/pilotless_leaderboard_pareto_front.csv};

        \addplot[myred, only marks, mark=star, mark size=4pt, ultra thick]
            coordinates {(1.235, 1.0207)};

  		\node[mylightgray2, labelstyle, inner sep=1.6mm] (d) at (1.25, 2.5) {Off-front};
  		\node[myblue, labelstyle] (d) at (0.75, 1.4) {Pareto front};
  		\node[myred, labelstyle] (d) at (1.02, 0.95) {Selected};
  		      
    \end{axis}
\end{tikzpicture}\label{fig:leaderboard_pareto}}
    \caption{Results of the \gls{aite} run for the pilotless receiver task.}
    \label{fig:leaderboard}
\end{figure}
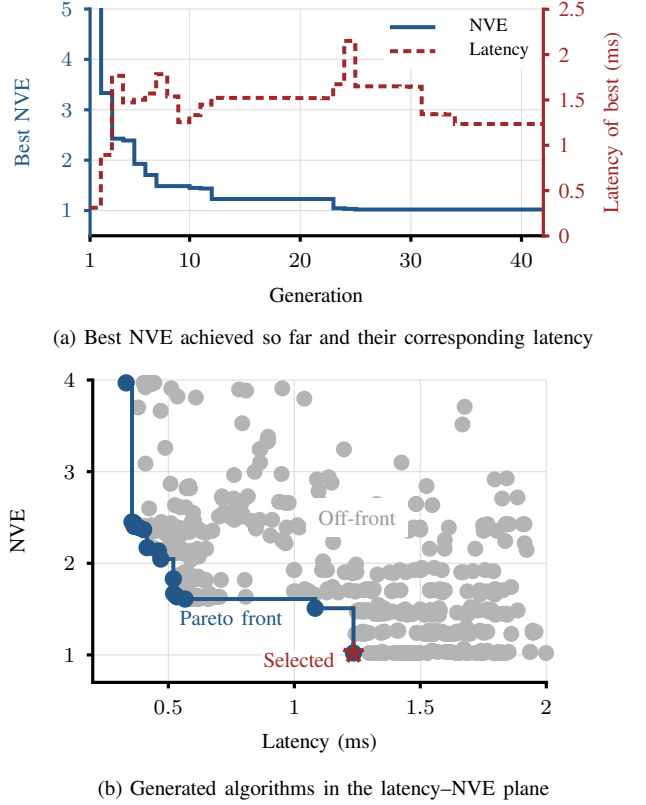

We ran the framework for \num{42} generations with GPT-5.5 as the \gls{llm} and \num{32} workers per generation on a cluster of \num{16} \glspl{gpu}.
Fig.~\ref{fig:leaderboard}\subref{fig:leaderboard_evolution} reports the evolution of the search.
The best \gls{nve} decreases rapidly during the first generations and reaches \(1.02\) at generation 25; among tied-best solutions, latency continues to decrease through generation 34.
In total, \gls{aite} produced \num{1240} functional receivers.  Rather than the monotone refinement of a single algorithm, the search can be described as a broad exploration that gradually concentrated around a small number of ideas.
The first generation instantiated more than a dozen different receiver families, most of them performing poorly.

Compared to the \gls{otfs} equalizer task, this task proved more challenging for \gls{aite}, requiring more than 20 generations to converge, as shown in Fig.~\ref{fig:leaderboard}\subref{fig:leaderboard_evolution}. Since no directly applicable prior art is available, \gls{aite} had to discover the key methods enabling blind detection, in what resembles a research process.
Progress originated from a small number of breakthroughs rather than from incremental tuning of a single approach.
The first key realization was that the structure of the learned constellation \(\Cc\) can be leveraged to initialize the channel estimate blindly.
Because the transmitted symbols are drawn from \(\Cc\) on every \gls{re}, the local empirical moments of the received grid relate to those of the constellation, i.e., over a small time--frequency neighborhood, \(\EE[Y^{m}] \approx H^{m}\,\EE[C^{m}]\) for integer \(m\), where \(C\) is uniformly distributed on \(\Cc\).
Estimating these moments and taking the \(m\)th root therefore yields a per-\gls{re} channel estimate up to a discrete \(m\)-fold phase ambiguity, with the residual ambiguity resolved by selecting the root that maximizes the likelihood \(\sum_{c \in \Cc} \exp{-\abs{Y - Hc}^{2}/N_{0}}\), which is possible only because the learned constellation, unlike \gls{qam}, has no rotational symmetry.
The resulting estimate is accurate enough to seed an \gls{em} refinement loop that alternates between soft symbol posteriors and a channel update.
This single idea moved the receiver out of the failure regime, and essentially every subsequent frontier solution descends from it.

A second breakthrough was not to commit to a single channel estimate, but to retain a small set of competing channel hypotheses per \gls{re} and use a log-sum-exp demapper that marginalizes the likelihood over both the candidate channels and the constellation points when computing the per-bit \glspl{llr}.
From these two ideas, the improvements that mattered were obtained by stacking complementary mechanisms onto this backbone rather than by replacing it.
Several of the families that had failed as standalone receivers reappeared at the Pareto front by being injected as refinements.

These auxiliary refinements are of three kinds. First, a belief-propagation refinement exploits the similarity of the channel gains at neighboring \glspl{re}.
These gains largely cancel in the ratio of the two received samples, leaving a relationship governed primarily by the two transmitted constellation points. Treating these pairwise relationships as soft constraints and propagating them across the grid sharpens the symbol decisions where the channel is hard to estimate directly. Second, a decision-directed channel-tracking candidate turns the receiver's soft estimate of the transmitted symbols into virtual pilots, from which an additional channel estimate is formed and smoothed. Third, a local affine model assumes that the channel varies approximately linearly within a small time--frequency patch and fits an affine plane to each patch to denoise the channel estimate.

In summary, after broad, divergent exploration, the lowest-\gls{nve} solutions share a common receiver backbone, with the bulk of the gain explained by a few enabling ideas and by the recombination of mechanisms that were individually insufficient.
Fig.~\ref{fig:leaderboard}\subref{fig:leaderboard_pareto} shows the Pareto front of the set of generated receivers together with the off-front solutions, with the view zoomed in on the low-\gls{nve} region.
Several algorithms attain the lowest \gls{nve} of \(1.02\). Among them, we select the one with the lowest latency, highlighted in Fig.~\ref{fig:leaderboard}\subref{fig:leaderboard_pareto}, for evaluation against the baselines.
A more detailed description of this algorithm is provided in Appendix~\ref{sec:pilotless-candidate-description}.

We benchmark the selected algorithm against two baselines: a pilot-based baseline (Fig.~\ref{fig:pilotless_systems}\subref{fig:pilotless_sys_a}) and the neural receiver trained jointly with the constellation for pilotless communication (Fig.~\ref{fig:pilotless_systems}\subref{fig:pilotless_sys_b}).
The pilot-based baseline follows a standard processing chain, i.e., least-squares channel estimation at the pilot positions \(\Pc\), linear interpolation of the channel estimates and error variances onto all data \glspl{re}, \gls{lmmse} equalization, and a posteriori probability demapping to \glspl{llr}.
This baseline is implemented with the corresponding components from Sionna~\cite{sionna} and employs a \(2^M\)-\gls{qam} constellation.

\begin{figure}[t]
    \centering
    \subfloat[Goodput on TDL-A (\gls{nlos})]{\begin{tikzpicture}
    \begin{axis}[standard_plotstyle,
        width=7cm, height=3.25cm,
        line width=1.5pt,
        xlabel={SNR (\unit{\decibel})}, ylabel={Goodput (\unit{\bit\per\resourceelement})},
        xmin=8, xmax=25, ymin=0, ymax=4.6,
        xtick={8,12,16,20,24},
        ytick={0,1,2,3,4},
        grid style={mylightgray4},
        legend columns=1,
        legend style={font=\scriptsize, column sep=1em,
                      draw=none, fill=white, fill opacity=0.75, text opacity=1,
                      row sep=-1pt, inner sep=1.5pt,
                      at={(0.97,0.09)}, anchor=south east},
        legend cell align=left,
    ]
        \addlegendimage{black}
        \addlegendentry{Perfect CSI}
        \addlegendimage{mpltab2, densely dashed, mark=*, mark size=1pt}
        \addlegendentry{Generated algorithm}
        \addlegendimage{mpltab1, mark=triangle*, mark size=1.4pt}
        \addlegendentry{Neural receiver}
        \addlegendimage{mpltab0, mark=square*, mark size=1pt}
        \addlegendentry{Pilot baseline}

        \addplot[black, forget plot]
            table[col sep=comma, x=snr, y=baseline_perfect_csi]{data/pilotless_goodput_tdla.csv};

        \addplot[mpltab1, mark=triangle*, mark size=1.4pt, forget plot]
            table[col sep=comma, x=snr, y=neural_receiver]{data/pilotless_goodput_tdla.csv};

        \addplot[mpltab2, densely dashed, mark=*, mark size=1pt, forget plot]
            table[col sep=comma, x=snr, y=agent_receiver]{data/pilotless_goodput_tdla.csv};

        \addplot[mpltab0, mark=square*, mark size=1pt, forget plot]
            table[col sep=comma, x=snr, y=baseline_ls]{data/pilotless_goodput_tdla.csv};
    \end{axis}
\end{tikzpicture}\label{fig:goodput_tdla}}\\[10pt]
    \subfloat[Goodput on TDL-D (\gls{los})]{\begin{tikzpicture}
    \begin{axis}[standard_plotstyle,
        width=7cm, height=3.25cm,
        line width=1.5pt,
        xlabel={SNR (\unit{\decibel})}, ylabel={Goodput (\unit{\bit\per\resourceelement})},
        xmin=8, xmax=25, ymin=0, ymax=4.6,
        xtick={8,12,16,20,24},
        ytick={0,1,2,3,4},
        grid style={mylightgray4},
    ]
        \addplot[black]
            table[col sep=comma, x=snr, y=baseline_perfect_csi]{data/pilotless_goodput_tdld.csv};

        \addplot[mpltab1, mark=triangle*, mark size=1.4pt]
            table[col sep=comma, x=snr, y=neural_receiver]{data/pilotless_goodput_tdld.csv};

        \addplot[mpltab2, densely dashed, mark=*, mark size=1pt]
            table[col sep=comma, x=snr, y=agent_receiver]{data/pilotless_goodput_tdld.csv};

        \addplot[mpltab0, mark=square*, mark size=1pt]
            table[col sep=comma, x=snr, y=baseline_ls]{data/pilotless_goodput_tdld.csv};
    \end{axis}
\end{tikzpicture}\label{fig:goodput_tdld}}
    \caption{Evaluation of the selected \gls{aite}-generated receiver algorithm against the pilot-based baseline and the neural receiver. The perfect-\gls{csi} upper bound uses \(2^M\)-\gls{qam} on every \gls{re}, exact channel coefficients, and the same demapper and \gls{ldpc} decoder as the pilot-based baseline.}
    \label{fig:eval_results}
\end{figure}
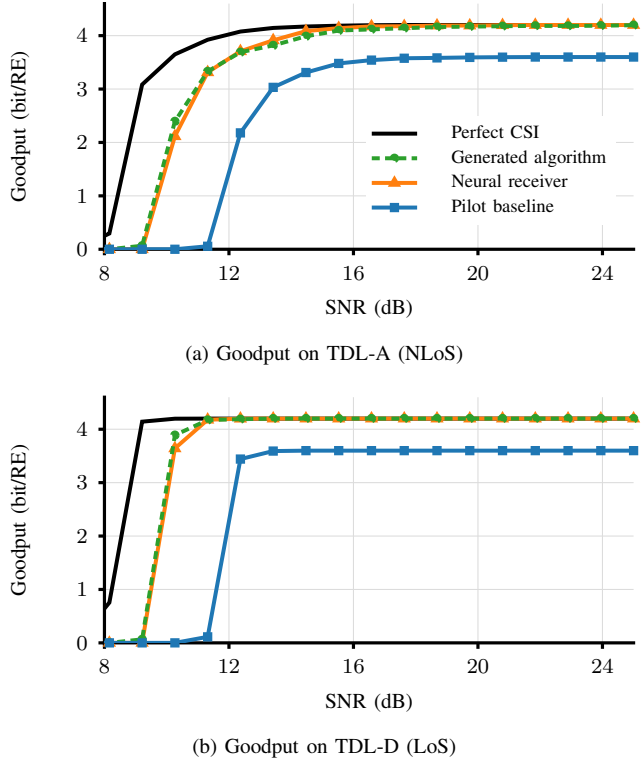

The goodput achieved by the evaluated receivers on TDL-A and TDL-D is reported in Fig.~\ref{fig:eval_results}\subref{fig:goodput_tdla} and \ref{fig:eval_results}\subref{fig:goodput_tdld}, respectively, and is defined as
\begin{equation}
    \text{Goodput} = \left(1 - \mathrm{BLER}\right) r_c\, r_p\, M,
    \label{eq:goodput}
\end{equation}
where \(r_c = 0.7\) is the code rate and \(M = 6\) is the number of bits per constellation symbol. The factor \(r_p\) is the fraction of \glspl{re} carrying data: \(r_p = 1\) for the pilotless approaches and the perfect-\gls{csi} upper bound, and \(r_p = 12/14\) for the pilot-based baseline.
As the figures show, the generated algorithm achieves goodput on par with that of the neural receiver on both channel models.
Both pilotless receivers realize the expected throughput gain over the pilot-based baseline, which comes from reallocating pilots to data.

The latency, measured on an NVIDIA~RTX~PRO~6000, is \qty{0.694}{\milli\second} for the neural receiver and \qty{1.23}{\milli\second} for the generated algorithm.
The longer latency is somewhat expected because, with the used PyTorch backend, the convolutional layers of the neural receiver map efficiently onto the \gls{gpu}, whereas the generated algorithm relies on non-neural signal-processing operations that are less amenable to straightforward parallelization.
Being explainable, however, the generated algorithm is easier to analyze and improve.
For instance, it could be extended to other modulation orders and multiple spatial streams, or accelerated through a dedicated CUDA implementation. Both are left to future work.

\section{Concluding Discussion}
\label{sec:conclusion}

Our experiments demonstrate that \gls{llm}-driven evolutionary search can discover novel wireless algorithms that are on par with or even surpass the current state of the art. Such methods allow for the parallel exploration and implementation of a large number of algorithmic ideas for a specific problem with the goal of building a performance--complexity Pareto front of solutions from which the most suitable can be selected. We believe that agentic tools like \gls{aite} are likely to become essential for the design of competitive algorithms in the future, and the broader impact of this development has been discussed by some of us~\cite{bjornson2026automating}. Developing \gls{aite} led us to identify several practical enablers and limitations, which we discuss in the remainder of this section.

\subsection{Impact of the Underlying LLM}
\label{subsec:llm_quality}

The quality of the generated algorithms is strongly affected by the capabilities of the \glspl{llm} used.
Experiments with smaller, less capable \glspl{llm} yielded significantly less competitive algorithms on complex tasks and required a larger number of iterations to converge.
Even when smaller \glspl{llm} proposed strong ideas, they often struggled to implement them correctly or to follow the instructions provided by the orchestrator, resulting in a higher rate of unexploitable algorithms.
This observation motivated the use of a frontier model (GPT-5.5) for the experiments reported in this article.
Each task took approximately four days on a cluster of \num{16} \glspl{gpu} and incurred an \gls{llm} inference cost of about USD~\num{2500} at the time of writing. However, this cost is likely to decrease significantly in the future. Quantifying the run-to-run variability of the search outcome remains open.

\subsection{The Importance of the Evaluation Tool}
\label{subsec:eval_bottleneck}

The definition of the evaluation metric is one of the most important design decisions because evolutionary search will relentlessly try to optimize for it and, as in supervised machine learning, there is a risk of overfitting to the evaluation tool (including channel models and target hardware platforms). One way to circumvent this could be a \emph{validation tool} whose outputs would only be used to detect when algorithms start to overfit so that they are not selected for refinement in future generations. In addition, we observed that \gls{aite} spends most of the running time evaluating candidate algorithms, as opposed to generating and implementing them. For the two tasks considered in this article, this evaluation is based on Monte Carlo simulations of \gls{bler} curves for the \gls{nve} as well as latency measurements of PyTorch-compiled code. The more accurate the desired metric estimates, the longer the evaluation takes. Designing an efficient evaluation tool is therefore a critical enabler for the scalability of evolutionary search methods. This becomes even more important when evaluations are carried out in large digital twin networks or with hardware-in-the-loop requiring time-consuming compilation or high-level synthesis.

\subsection{Hyperparameter Tuning}
\label{subsec:hp_finetuning_discussion}

Most generated algorithms have hyperparameters, whose number typically grows as the evolutionary search progresses and the algorithms become more complex. As described in Section~\ref{subsubsec:hp_finetuning}, \gls{aite} provides a hyperparameter fine-tuning stage, which was applied to every candidate algorithm in the experiments reported here.
However, Bayesian optimization becomes impractical with even a few dozen hyperparameters.
One solution would be to replace Bayesian optimization with gradient descent for differentiable continuous hyperparameters, combined with a separate optimization method for those defined in discrete spaces, such as the number of iterations.
A different approach would be to run substantially fewer generations of the framework and to perform more exhaustive hyperparameter fine-tuning for each candidate.
The optimal allocation problem remains unresolved.

\subsection{Idea Generation: Exploration Versus Exploitation}
\label{subsec:exploration_exploitation}

Evolutionary search frameworks like \gls{aite} need to solve an exploration--exploitation tradeoff at the start of each generation, when a finite number of workers must be tasked either with exploring novel ideas or refining ideas from previous generations. Insufficient exploration may cause potential breakthrough ideas to be missed, while insufficient exploitation may cause promising ideas to be discarded prematurely. In \gls{aite}, the orchestrator \gls{llm} is responsible for solving this tradeoff through careful prompt refinement. Other frameworks adopt different strategies~\cite{cemri2026adaevolve,lange2025shinka, skydiscover}. Improving the sample efficiency of the search is an important direction for future work that would also allow reallocating resources to hyperparameter tuning, cf.~Section~\ref{subsec:hp_finetuning_discussion}.

\subsection{Measuring Complexity}
\label{subsec:complexity_measurement}

We can characterize algorithmic complexity in many ways. Theoretical complexity concepts such as floating-point operations or asymptotic analysis are important for high-level comparisons, but they are often unreliable indicators of actual latency or throughput on target hardware, such as a \gls{gpu} or a central processing unit. Unfortunately, measuring these concrete metrics requires direct access to the target hardware, which is often infeasible in large compute clusters. Frameworks like \gls{aite} can in principle use arbitrary complexity metrics, but, much like the evaluation metric, these metrics must be chosen carefully to properly steer the search toward practically relevant outcomes.

\appendices
\section{OTFS Input--Output Relation}
\label{sec:otfs-io}

Based on the system model in Section~\ref{sec:otfs-system-model}, we derive a closed-form input--output relation for the \gls{otfs} system for use in the evaluation tool's Monte Carlo simulations.

Let us introduce the normalized \gls{dd} coordinates
\begin{equation}
\widetilde{l}_i \triangleq \tau_i M\Df = l_i+\varepsilon_i,
\quad
\widetilde{k}_i \triangleq \nu_i N (T+\Tcp) = k_i+\kappa_i,
\label{eq:norm-coords}
\end{equation}
with nearest integers $l_i,k_i\in\ZZ$ and fractional parts
$\varepsilon_i, \kappa_i\in(-\tfrac12,\tfrac12]$. The received signal $y[l,k]$ on the \gls{dd} grid after matched filtering, sampling at rate
$1/\Ts$, \gls{cp} removal, and the unitary per-sub-block $M$-point \gls{dft} and symplectic finite Fourier transform
is then given by \eqref{eq:io}, where
\begin{align}
    \Phi_i(l) &= e^{-j2\pi\nu_i\tau_i}\,e^{\,j2\pi\nu_i l\Ts},\\
    \Dir_Q(\xi) &= \frac{1}{Q}\sum_{q=0}^{Q-1}e^{\,j2\pi q\xi/Q}
\end{align}
are per-path phase terms and $Q$-point Dirichlet kernels, respectively. Here, $(\cdot)_Q$
denotes reduction modulo $Q$, and $w[l,k]$ is standard complex \gls{awgn} with
variance $N_0$. The values of $\gamma_\tau$ and $\gamma_\nu$ run over the
windows 
\begin{align}
    \Wt &= \{-\lfloor M/2\rfloor,\dots,\lceil M/2\rceil-1\},\\
    \Wn &= \{-\lfloor N/2\rfloor,\dots,\lceil N/2\rceil-1\}
\end{align}
of size exactly $M$ and $N$ for either parity.

Collecting the \gls{dd} symbols into vectors $\mathbf x,\mathbf y,\mathbf
w\in\CC^{MN}$ by assigning each cell $(l,k)$ the scalar index
$n(l,k)\triangleq k+Nl$ (row-by-row stacking, with the Doppler index varying fastest, so that
$[\mathbf x]_{n(l,k)}=x[l,k]$, and likewise for $\mathbf y$ and $\mathbf w$), we can
write the input--output relation \eqref{eq:io} compactly as
\begin{equation*}
    \mathbf y = \mathbf H\,\mathbf x + \mathbf w,
\end{equation*}
where $\mathbf H\in\CC^{MN\times MN}$ is the effective \gls{dd} channel matrix with elements
given by \eqref{eq:Hentry}. For the full windows $\Wt,\Wn$, the summations over the Kronecker deltas have exactly one nonzero term for every $l,l',k,k'$, so that $\mathbf H$ is generally dense for noninteger delays or Dopplers.

\begin{figure*}[!t]
    \begin{equation} \label{eq:io}
        y[l,k] = \sum_{i=1}^{P} h_i\,\Phi_i(l)
        \sum_{\gamma_\tau \in \Wt}\sum_{\gamma_\nu \in \Wn}
        \Dir_M(\gamma_\tau-\varepsilon_i)
        \Dir_N(\gamma_\nu+\kappa_i)\,
        x\!\left[\modM{l-l_i-\gamma_\tau},\,\modN{k-k_i+\gamma_\nu}\right] + w[l,k]
    \end{equation}
    \hrulefill
    \end{figure*}

\begin{figure*}[!t]
    \begin{equation} \label{eq:Hentry}
        [\mathbf H]_{n(l,k),\,n(l',k')} = \sum_{i=1}^{P} h_i\,\Phi_i(l)
        \sum_{\gamma_\tau \in \Wt}
        \Dir_M(\gamma_\tau-\varepsilon_i)
        \delta\!\left[l'-\modM{l-l_i-\gamma_\tau}\right]
        \sum_{\gamma_\nu \in \Wn}
        \Dir_N(\gamma_\nu+\kappa_i)\,
        \delta\!\left[k'-\modN{k-k_i+\gamma_\nu}\right]
    \end{equation}
    \hrulefill
    \end{figure*}

In practice, the 
windows can be truncated with little loss in accuracy to $\Wtt=\{-N_\tau,\dots,N_\tau\}$ and $\Wnt=\{-N_\nu,\dots,N_\nu\}$ with
$2N_\tau{+}1\ll M$ and $2N_\nu{+}1\ll N$. When these truncated windows are used in \eqref{eq:Hentry},
only the $(2N_\tau+1)(2N_\nu+1)$ columns nearest the
integer shift $(l-l_i,k-k_i)$ are nonzero for each path, yielding a row-sparse matrix $\widetilde{\mathbf H}$ with at
most $S=P\,(2N_\tau+1)(2N_\nu+1)$ nonzero entries per row.

Alternatively, rather than fixing the windows a priori, one can compute the
exact dense $\mathbf H$ and retain in each row only the \topk\ entries
with the largest absolute value. This yields a controlled sparse
approximation whose accuracy increases monotonically with $k$, and---in contrast
to the fixed window above---adapts to the actual location of the dominant taps,
which is advantageous under pronounced fractional-delay or fractional-Doppler
leakage. Equalizers can choose to work with such truncated channel matrices for efficiency, whereas the simulation of $\yv$ always uses the full channel matrix.

\section{Discussion of an OTFS Equalizer Candidate}
\label{sec:otfs-candidate-description}

For an observation $\mathbf{y}$ and perfect channel knowledge $\mathbf{H}$,
the equalizer's goal is to compute the posterior marginals $p(x_r|\mathbf{y})$ for
$r\in\{0,\ldots,MN-1\}$ under the channel likelihood \eqref{eq:matrix} and a discrete
prior $p(x_r)$, which accounts for the fact that all symbols are drawn from a \gls{qam} constellation
$\mathcal{Q}=\{\alpha_1,\ldots,\alpha_Q\}$ of order $Q$. Computing the joint
posterior $p(\mathbf{x}|\mathbf{y}) \propto p(\mathbf{y}|\mathbf{x})p(\mathbf{x})$
is prohibitive because all symbols are discrete and coupled by the matrix $\mathbf{H}$.
The algorithm sidesteps this by approximating the posterior
$p(\mathbf{x}|\mathbf{y})$ with a Gaussian $q(\mathbf{x})$
whose marginal moments are made consistent between two simpler views of the
problem, iterating until both agree. This two-module moment-matching
construction is an \gls{ec} approximation~\cite{opper2005ec} and works as
follows:

\subsection{Module~A: Linear Model with Gaussian Prior}
The first view corresponds to the linear model
\eqref{eq:matrix} with a Gaussian symbol prior $p_B(\mathbf{x})$ which it
receives from Module~B. This prior is defined by its natural parameters, the precision $\mathbf{p}_B \in
\mathbb{R}^{MN}_+$ and precision-weighted mean $\boldsymbol{\eta}_B \in
\mathbb{C}^{MN}$. In the first iteration this message is initialized to the unit-energy Gaussian prior,
$\mathbf{p}_B = \onev$ and $\boldsymbol{\eta}_B = \mathbf{0}$, so that
Module~A reduces to a plain linear (\gls{lmmse}-type) estimate. Under this
assumption, the corresponding posterior is also
Gaussian with natural parameters $\tilde{\mathbf{p}}_A$ and
$\tilde{\boldsymbol{\eta}}_A$:
\begin{align}
     \tilde{\mathbf{p}}_A  &= \onev \oslash \operatorname{diagpart} \left( \mathbf{A}^{-1} \right) \label{eq:p_tilde_a}\\
    \tilde{\boldsymbol{\eta}}_A &= \tilde{\mathbf{p}}_A \odot \left[\mathbf{A}^{-1} \left( \boldsymbol{\eta}_B + \frac{1}{N_0} \mathbf{H}^\mathsf{H}\mathbf{y}\right)\right] \label{eq:eta_tilde_a}\\
    \mathbf{A} &= \operatorname{diag}\left(\mathbf{p}_B \right) + \frac{1}{N_0} \mathbf{H}^\mathsf{H}\mathbf{H},
\end{align}
where $\odot$ and $\oslash$ denote elementwise multiplication and division, respectively.
Module~A sends to Module~B only the extrinsic Gaussian with natural parameters
\begin{equation}
    \mathbf{p}_A = \tilde{\mathbf{p}}_A - \mathbf{p}_B,\quad
    \boldsymbol{\eta}_A = \tilde{\boldsymbol{\eta}}_A - \boldsymbol{\eta}_B.
    \label{eq:extrinsic_a}
\end{equation} 

\subsection{Module~B: QAM Denoiser via Moment Matching}
This view treats the discrete prior exactly
and accounts for the channel only through the message it receives from Module~A,
which it
interprets as a per-symbol Gaussian pseudo-observation of the transmitted
symbols
\begin{equation}
    \mathbf{r}_A = \mathbf{x} + \mathbf{n},
\end{equation}
where $\mathbf{r}_A=\boldsymbol{\eta}_A \oslash \mathbf{p}_A$, and
$\mathbf{n}\sim\CN(0, \mathop{\text{diag}}\left(\mathbf{p}_A\right)^{-1})$.
Assuming equiprobable constellation symbols, it then computes for each symbol index $r$ the discrete posterior $\gamma_{r,q} = p(x_r=\alpha_q| r_{A,r})$ over the $Q$ constellation
points:
\begin{equation}
    \gamma_{r,q} = \frac{\exp{-p_{A,r}\,|r_{A,r}-\alpha_q|^2}}
                        {\sum_{q'=1}^{Q}\exp{-p_{A,r}\,|r_{A,r}-\alpha_{q'}|^2}}.
    \label{eq:gamma}
\end{equation}
As this tilted posterior is not Gaussian, Module~B projects it back onto a
Gaussian by matching its first two moments, giving the posterior mean
$\tilde{\mathbf{r}}_B$ and variance $\tilde{\mathbf{v}}_B$ with
\begin{equation}
   \tilde{r}_{B,r} = \sum_{q=1}^{Q} \gamma_{r,q}\,\alpha_q, \quad
    \tilde{v}_{B,r} = \sum_{q=1}^{Q} \gamma_{r,q}\,|\alpha_q|^2
        - \left| \tilde{r}_{B,r} \right|^2,
\end{equation}
which lead to the natural parameters $\tilde{\mathbf{p}}_B=\onev\oslash\tilde{\mathbf{v}}_B$ and $\tilde{\boldsymbol{\eta}}_B=\tilde{\mathbf{p}}_B \odot \tilde{\mathbf{r}}_B$.
Module~B then forms the extrinsic message
\begin{equation}
    \mathbf{p}_B^{\mathrm{ext}} = \tilde{\mathbf{p}}_B - \mathbf{p}_A,\quad
    \boldsymbol{\eta}_B^{\mathrm{ext}} = \tilde{\boldsymbol{\eta}}_B - \boldsymbol{\eta}_A
    \label{eq:extrinsic_b}
\end{equation}
and sends a damped version back to Module~A:
\begin{equation}
    \mathbf{p}_B \leftarrow (1-\beta)\,\mathbf{p}_B + \beta\,\mathbf{p}_B^{\mathrm{ext}},\quad
    \boldsymbol{\eta}_B \leftarrow (1-\beta)\,\boldsymbol{\eta}_B + \beta\,\boldsymbol{\eta}_B^{\mathrm{ext}},
    \label{eq:damping}
\end{equation}
with $\beta\in(0,1]$.
This exchange
is repeated until the marginals of the two modules agree. In this generic form,
the per-bit \glspl{llr} are read off from the final symbol posteriors
$\gamma_{r,q}$.

\subsection{Optimized Implementation}
The generated algorithm is a performance- and latency-optimized version of this
algorithm based on the following key modifications:

\begin{enumerate}
  \item \textbf{Per-Doppler block-diagonalization.}
  The input--output relation \eqref{eq:io} is a
  block-circulant convolution along the Doppler index which can be
  diagonalized by an $N$-point \gls{dft} along the Doppler
  axis~\cite{raviteja2018}. This decouples the $MN\times MN$ system into $N$ independent $M\times M$
  blocks which can be treated in parallel. The implementation further shares a
  single precision per delay across all Doppler bins,
  $\mathbf{p}_B\in\RR_+^{M}$ (the mean $\boldsymbol{\eta}_B\in\CC^{M\times N}$ stays
  per cell), so the same $\operatorname{diag}(\mathbf{p}_B)$ regularizes every block.

\item \textbf{No matrix inversion.} Module~A never forms $\mathbf{A}^{-1}$. In
\eqref{eq:eta_tilde_a}, Module~A uses Cholesky factorization followed by forward/back substitution to obtain $\tilde{\boldsymbol{\eta}}_A$ for each Doppler block. The posterior precisions in
\eqref{eq:p_tilde_a} are approximated by a second-order Jacobi estimate of
$\operatorname{diagpart}(\mathbf{A}^{-1})$, averaged over
Doppler to one variance per delay index.

\item \textbf{Few iterations plus one Richardson half-step.}
The algorithm runs for only two full \gls{ec} iterations. In the third, the
Cholesky solve to compute  \eqref{eq:eta_tilde_a} is replaced by a single low-complexity
damped Jacobi (i.e., diagonally preconditioned Richardson) update~\cite{saad2003iterative}:
\begin{equation*}
  \mathbf{x}_{\mathrm{R}} = \tilde{\mathbf{r}}_B
    + \omega\,\operatorname{diag}(\mathbf{A})^{-1}
      \left( \boldsymbol{\eta}_B + \tfrac{1}{N_0}\mathbf{H}^\mathsf{H}\mathbf{y}
             - \mathbf{A}\,\tilde{\mathbf{r}}_B \right),
\end{equation*}
with relaxation factor $\omega$. The associated precision $\tilde{\mathbf{p}}_A$ is obtained from \eqref{eq:p_tilde_a}, and the resulting \emph{Richardson cavity}
follows by the same extrinsic subtraction as in \eqref{eq:extrinsic_a}, now with
$\mathbf{x}_{\mathrm{R}}$ in place of the exact posterior mean,
\begin{equation}
  \mathbf{p}_A^{\mathrm{R}} = \tilde{\mathbf{p}}_A - \mathbf{p}_B,\quad
  \mathbf{r}_A^{\mathrm{R}} = \big(\tilde{\mathbf{p}}_A \odot \mathbf{x}_{\mathrm{R}}
        - \boldsymbol{\eta}_B\big) \oslash \mathbf{p}_A^{\mathrm{R}}.
\end{equation}
This cavity is not fed back to Module~B; it is passed directly to the final
demapper.

\item \textbf{Fused demapping from multiple log-metrics.}
    Let $\mathcal{L}_{r,q}(\mathbf{r},\mathbf{p}) = -p_r\,|r_r-\alpha_q|^2$ denote
    the constellation log-metric of a Gaussian pseudo-observation
    $(\mathbf{r},\mathbf{p})$, i.e., the numerator exponent of $\gamma_{r,q}$ in
    \eqref{eq:gamma}. Rather than demapping a single posterior, the
    implementation linearly fuses three such metrics, evaluated at the
    Richardson cavity $(\mathbf{r}_A^{\mathrm{R}},\mathbf{p}_A^{\mathrm{R}})$, the
    last cavity $(\mathbf{r}_A,\mathbf{p}_A)$, and the last linear posterior
    $(\tilde{\boldsymbol{\eta}}_A\oslash\tilde{\mathbf{p}}_A,\tilde{\mathbf{p}}_A)$,
    \begin{align}
      \mathcal{L}^{\mathrm{f}}_{r,q} &=
        w_c\,\mathcal{L}_{r,q}(\mathbf{r}_A^{\mathrm{R}},\mathbf{p}_A^{\mathrm{R}})
      + w_t\,\mathcal{L}_{r,q}(\mathbf{r}_A,\mathbf{p}_A) \nonumber\\
      &
      + w_\ell\,\mathcal{L}_{r,q}(\tilde{\boldsymbol{\eta}}_A\oslash\tilde{\mathbf{p}}_A,\tilde{\mathbf{p}}_A)
      \label{eq:fusion}
    \end{align}
    and reads the per-bit \glspl{llr} by max-log demapping,
    \begin{equation}
      \lambda_{r,k} = \max_{q\in\mathcal{S}_k^{1}}\mathcal{L}^{\mathrm{f}}_{r,q}
                    - \max_{q\in\mathcal{S}_k^{0}}\mathcal{L}^{\mathrm{f}}_{r,q},
      \label{eq:fused_llr}
    \end{equation}
    where $\mathcal{S}_k^{b}=\{q:\text{the }k\text{th label bit of }\alpha_q\text{ is }b\}$.

\item \textbf{Scalings and precomputation.}
The noise variance is scaled, $\tilde N_0 = s_{N_0} N_0$, to correct for model
mismatch and used in place of $N_0$. The data terms
$\tilde N_0^{-1}\mathbf{H}^\mathsf{H}\mathbf{H}$ and
$\tilde N_0^{-1}\mathbf{H}^\mathsf{H}\mathbf{y}$ are formed once per frame and
reused across iterations, leaving only $\operatorname{diag}(\mathbf{p}_B)$ and the
prior-pull term in \eqref{eq:eta_tilde_a} to update. The output \glspl{llr} are
finally rescaled by a factor $s_\lambda$ before the \gls{ldpc} decoder.

\item \textbf{Hyperparameters.} The tuned hyperparameters comprise the channel sparsity, numerical clipping thresholds, relaxation and damping factors, fusion weights, noise and \gls{llr} scalings, and final demapping rule. They are selected by the framework's postrun hyperparameter optimization (Section~\ref{subsubsec:hp_finetuning}).
\end{enumerate}

\section{Discussion of a Pilotless OFDM Receiver Candidate}
\label{sec:pilotless-candidate-description}
 
The receiver maps the received grid \(\Ym\) to \glspl{llr} through a sequence of stages.
For clarity, we only describe the key operation of each stage and omit details such as windowing, damping, and reliability gating.
The complete implementation is available in the code repository~\cite{aite2026repo}.
Throughout, \(\langle\cdot\rangle_{f,t}\) denotes a local average over a small time--frequency window centered at \((f,t)\).

\subsection{Multiscale Moment Initialization}
The receiver first forms a set of candidate channel estimates using moment matching.
Since the constellation has unit average power, the amplitude follows from the received energy:
\begin{equation}
    \widehat{|H|}_{f,t} = \sqrt{\max\!\big(\langle |Y|^2\rangle_{f,t} - N_0,\, 0\big)}.
    \label{eq:gen_amp}
\end{equation}
The phase is recovered from the \(m\)th-order moments through the approximation \(\langle Y^m\rangle_{f,t}\approx H_{f,t}^m\,\mu_m\), with \(\mu_m=\frac{1}{\abs{\Cc}}\sum_{c\in\Cc} c^m\).
Taking the \(m\)th root leaves an \(m\)-fold ambiguity,
\begin{equation}
    \widehat\theta^{(m,r)}_{f,t} = \frac{1}{m}\arg\!\left(\frac{\langle Y^m\rangle_{f,t}}{\mu_m}\right) + \frac{2\pi r}{m}, \quad r=0,\dots,m-1,
    \label{eq:gen_phase}
\end{equation}
so that the orders \(m\in\{2,3,4,6\}\) used in the algorithm yield \(2+3+4+6=15\) candidate phases, hence \(15\) candidate channels \(\widehat H^{(m,r)}_{f,t}=\widehat{|H|}_{f,t}\,e^{j\widehat\theta^{(m,r)}_{f,t}}\).
The moments are estimated at three time--frequency window aspect ratios for the averaging \(\langle\cdot\rangle_{f,t}\) to cover different Doppler and delay profiles.
At each scale, the $N_{\text{top}}$ candidates with the highest log-likelihood,
\begin{equation}
    S_{f,t}^{(m,r)} =
    \left\langle \ln{\sum_{c\in\Cc} \exp{ -\frac{ \abs{Y_{f,t} - \widehat{H}^{(m,r)}_{f,t} c}^2 }{N_0}}} \right\rangle_{f,t},
    \label{eq:gen_score}
\end{equation}
are retained, leaving \(3 N_{\text{top}}\) winners per \gls{re}.
To these are added \(N_{\text{top}}\) candidates obtained by fusing the rank-matched winners of the three scales through a score-weighted average, leading to a cross-scale blended estimate where the scales agree, while the per-scale winners are retained where they disagree.
The result is a pool of \(K = 4 N_{\text{top}}\) candidates per \gls{re}, which we reindex by a single label \(k\) and denote by \(\{\widehat H^{(k)}_{f,t}\}_{k=1}^{K}\), with corresponding log-likelihoods \(\{S^{(k)}_{f,t}\}_{k=1}^{K}\).

\subsection{Hypothesis Smoothing}
Since the moment-based estimates are computed locally and the channel varies smoothly, the discrete choice among the \(K\) candidates should be coherent across the grid.
This is achieved by modeling the resource grid as a conditional random field over the candidates.
Let us denote by $\ell_u$ the random variable that assigns to each \gls{re} $u$ the ``true'' channel candidate index out of the $K$ candidates, and by $\boldsymbol\ell=\{\ell_u\}$ the grid-wide assignment.
The posterior on $\boldsymbol\ell$ is modeled as
\begin{multline}
    P(\boldsymbol\ell\mid\Ym) \propto \mathop{\mathrm{exp}}\Bigg( -\tfrac1\tau\Big( -\alpha\!\sum_{u} S^{(\ell_{u})}_{u}\\
    + \lambda\!\!\sum_{(u,v)\in\Ec}\!\! w_{uv}\abs{\widehat H^{(\ell_u)}_u - \widehat H^{(\ell_v)}_v}^2 \Big) \Bigg),
    \label{eq:gen_crf_post}
\end{multline}
where $\alpha$ and $\lambda$ are hyperparameters and $\Ec$ collects pairs of neighboring \glspl{re}.
The $w_{uv}$ are uniform weights over a boxcar window in time and frequency.
Since the scores $S^{(k)}_{u}$ are log-likelihoods, the unary term is the log-likelihood of $\boldsymbol\ell$ over the resource grid under an independence assumption. The pairwise term penalizes neighbors that select channels with dissimilar values.

Computing the marginals of \eqref{eq:gen_crf_post} is intractable, so the posterior is approximated by a product of per-node distributions \(Q(\boldsymbol\ell)=\prod_{u} q_{u}(\ell_{u})\), with \(q_{u}(k)\ge 0\), \(\sum_{k=1}^{K} q_{u}(k)=1\), chosen to minimize the Kullback--Leibler divergence \(\mathrm{KL}\!\big(Q\,\|\,P(\cdot\mid\Ym)\big)\).
Each factor \(q_{u}(k)\) is then the approximate marginal probability that node \(u\) takes candidate \(k\).
Zeroing the gradient of this divergence with respect to each \(q_{u}(k)\) yields the fixed-point update
\begin{equation}
    q_{u}(k) \propto \mathop{\mathrm{exp}} \Bigg( \tfrac{1}{\tau}\Big(\alpha\,S^{(k)}_{u} - \lambda\,\abs{\widehat H^{(k)}_{u}-\bar H_{u}}^2\Big) \Bigg)
    \label{eq:gen_crf}
\end{equation}
for all \(k\in\{1,\dots,K\}\), in which each node sees its neighbors only through
\begin{equation}
    \bar{H}_{u} = \sum_{v \in \Nc(u)} w_{uv} \sum_{k=1}^{K} q_{v}(k)\, \widehat{H}^{(k)}_{v},
    \label{eq:gen_hbar}
\end{equation}
that is, the soft-assigned average channel of neighbors $\Nc(u)$.
Note that the implementation approximates~\eqref{eq:gen_hbar} by a boxcar window centered on $u$, which also includes the node's own soft channel.
Since \(\bar H_{u}\) in \eqref{eq:gen_hbar} depends on the same assignments updated by \eqref{eq:gen_crf}, the two are applied alternately: starting from the unary-only assignment \(q_{u}(k)\propto\exp{\alpha\,S^{(k)}_{u}/\tau}\), a few sweeps iterate \eqref{eq:gen_hbar} and~\eqref{eq:gen_crf}.

\subsection{Expectation-Maximization}
The previous two stages aim to provide strong initializations for an \gls{em} loop that jointly refines the channel estimates and computes posterior symbol probabilities.
Each of the \(K\) candidates is refined in parallel, with \(\Ym\) as the observed data, the channel \(\Hm\) as the parameter to be estimated, and the transmitted symbols as the latent variables.
The E-step computes soft symbol posteriors under the current channel,
\begin{equation}
    p^{(k)}_{f,t}(c) \propto \exp{-\abs{Y_{f,t}-\widehat H^{(k)}_{f,t}\,c}^2/N_0}, \quad c\in\Cc.
    \label{eq:gen_estep}
\end{equation}
The M-step re-estimates the channel by treating the resulting soft symbols as pseudo-pilots, implementing a confidence-weighted least-squares fit over the local window,
\begin{equation}
    \widehat H^{(k)}_{f,t} \leftarrow \frac{\langle w^{(k)}_{f,t}\,Y_{f,t}\,\widehat c^{(k)*}_{f,t}\rangle_{f,t}}{\langle w^{(k)}_{f,t}\,\widehat v^{(k)}_{f,t}\rangle_{f,t}},
    \label{eq:gen_mstep}
\end{equation}
with soft moments \(\widehat c^{(k)}_{f,t}=\sum_{c}p^{(k)}_{f,t}(c)\,c\), \(\widehat v^{(k)}_{f,t}=\sum_{c}p^{(k)}_{f,t}(c)\,\abs{c}^2\), and a per-candidate weight \(w^{(k)}_{f,t}\propto (q_{f,t}(k))^{\kappa}\,(\max_c p^{(k)}_{f,t}(c))^{\beta}\) that combines the posterior \(q_{f,t}(k)\) from the previous stage with the symbol-decision probabilities, where $\kappa$ and $\beta$ are hyperparameters.

After the \gls{em} loop, a probability mass function over the candidates is defined by blending the smoothed posterior with how well each candidate explains the received data,
\begin{equation}
    \pi_{f,t}(k)\propto\big(q_{f,t}(k)\big)^{\rho}\,\exp{\tfrac{1-\rho}{\tau_\pi}\,S^{(k)}_{f,t}}, \quad 1 \leq k \leq K,
    \label{eq:gen_trust}
\end{equation}
such that $\sum_{k=1}^{K}\pi_{f,t}(k)=1$, where \(S^{(k)}_{f,t}\) is the local mixture log-likelihood of the form~\eqref{eq:gen_score} re-evaluated at the converged channel \(\widehat H^{(k)}_{f,t}\), \(q_{f,t}(k)\) is the mean-field posterior obtained by reapplying the smoothing of~\eqref{eq:gen_crf} to these refined scores, and \(\rho\in[0,1]\) and \(\tau_\pi\) are hyperparameters.
These weights yield a fused posterior on the transmitted symbols
\begin{equation}
    \bar p_{f,t}(c)=\sum_{k=1}^{K}\pi_{f,t}(k)\,p^{(k)}_{f,t}(c), \quad c\in\Cc.
    \label{eq:gen_fused_post}
\end{equation}
The loop also returns a per-\gls{re} channel-estimation variance \(\sigma^2_{H,f,t}\), used later for computing the \glspl{llr}.

\subsection{Ratio-Prior Belief Propagation}

While \eqref{eq:gen_fused_post} gives probability mass functions for all symbols in the resource grid, the true posteriors are coupled, because the channel is approximately constant over neighboring \glspl{re}. Hence, more accurate marginals of the grid-wide posterior can be obtained through a belief propagation-inspired scheme on a Markov random field.
This first refinement approximates, for each \gls{re}, the marginal of its symbol under the grid-wide posterior,
\begin{equation}
    P(c_{f,t}\mid\Ym)=\!\!\sum_{\boldsymbol c\,\setminus\, c_{f,t}}\!\! P(\boldsymbol c\mid\Ym),
    \label{eq:gen_ratio_marginal}
\end{equation}
where \(\boldsymbol c=\{c_{f,t}\}\) is the grid-wide symbol labeling, and turns it into a per-\gls{re} symbol log-domain refinement \(\Gamma_{f,t}(c)\) that is fed back into the estimation.
As all the \glspl{re} are coupled through the unknown channel,~\eqref{eq:gen_ratio_marginal} is intractable.

It is assumed that the dependency between symbols is restricted to time- and frequency-adjacent \glspl{re}, which makes \(P(\boldsymbol c\mid\Ym)\) a pairwise Markov random field on the resource grid.
For such a field, the Hammersley--Clifford theorem gives the factorization into unary and pairwise potentials,
\begin{equation}
    P(\boldsymbol c\mid\Ym)\;\propto\;\prod_{u}\psi_u(c_u)\!\!\prod_{(u,v)\in\Ec}\!\!\psi_{uv}(c_u,c_v),
    \label{eq:gen_ratio_hc}
\end{equation}
where $u$ and $v$ index \glspl{re} and \(\Ec\) is the set of neighboring \gls{re} pairs.
The two potentials are set as follows.
The unary potential is set to \(\psi_u(c)=\bar p_u(c)^{\,w_{\mathrm{src}}}\), where \(w_{\mathrm{src}}\) is a hyperparameter.
The pairwise potential scores adjacent symbols through the ratio of their received samples: two adjacent \glspl{re} are assumed to experience approximately the same channel (\(H_u\approx H_v\)), so the channel approximately cancels in that ratio,
\begin{equation}
    \frac{Y_u}{Y_v}=\frac{H_u X_u + W_u}{H_v X_v + W_v}\approx \frac{X_u}{X_v},
    \label{eq:gen_ratio}
\end{equation}
and the observed ratio \(q_{uv}=Y_u/Y_v\) depends on the two symbols only.
The corresponding potential is
\begin{equation}
    \begin{aligned}
        \psi_{uv}(c_u,c_v) &= e^{\phi_{uv}(c_u,c_v)},~\text{with}\\
        \phi_{uv}(c_u,c_v) &= -g\,\ln{1+\frac{\abs{q_{uv}-c_u/c_v}^2}{\sigma^2_{uv}}}
    \end{aligned},
    \label{eq:gen_ratio_phi}
\end{equation}
with \(g\) a hyperparameter and ratio-error variance \(\sigma^2_{uv}\propto N_0(1+\abs{q_{uv}}^2)/\abs{Y_v}^2\).
This factor has heavy Cauchy-like tails, which suppress edges near a deep fade where \(\abs{Y_v}\) is small and the ratio is unreliable.

The marginals~\eqref{eq:gen_ratio_marginal} of the field~\eqref{eq:gen_ratio_hc} are approximated by a belief propagation-inspired approach, with each symbol restricted to the set \(\Mc_{f,t}\subset\Cc\) of its most likely values under the \gls{em} posterior.
The edge messages are
\begin{equation}
    \mu_{v\to u}(c) \propto \sum_{c'\in\Mc_v}\exp{\phi_{uv}(c,c') + w_{\mathrm{bp}}\,b_v(c')},
\end{equation}
with \(w_{\mathrm{bp}}\) a hyperparameter.
The node belief $b_u(c)$ combines the incoming log-messages with the unary log-potential,
\begin{multline}
    b_u(c) =\frac{1}{d_u}\sum_{v\in\Nc(u)} w_{uv}\,\ln{\mu_{v\to u}(c)}\\+ w_{\mathrm{src}}\,\Rc_u\,\ln{\bar p_u(c)},
\end{multline}
where \(\Nc(u)\) is the set of neighbors of \(u\) and \(d_u=\sum_{v\in\Nc(u)} w_{uv}\) is the weighted node degree.
Here \(\Rc_u\in[0,1]\) is a per-node reliability and \(w_{uv}\) an edge weight, whose detailed definitions are omitted.
While the messages and beliefs could be updated alternately over the grid, the implementation runs a single sweep, comprising one frequency- and one time-adjacent pass.

The resulting beliefs, mapped back onto the constellation, give the per-\gls{re} symbol refinement \(\Gamma_{f,t}(c)\).
This refinement is added as a symbol log-prior in a further \gls{em} pass, which re-estimates the channel \(\widehat H^{(k)}_{f,t}\) and the symbol posteriors using the following E-step in place of~\eqref{eq:gen_estep}:
\begin{equation}
    p^{(k)}_{f,t}(c) \propto \exp{-\abs{Y_{f,t}-\widehat H^{(k)}_{f,t}\,c}^2/N_0 + w'_{\Gamma}\,\Gamma_{f,t}(c)},
    \label{eq:gen_estep_prior}
\end{equation}
where \(w'_{\Gamma}\) is a hyperparameter.

\subsection{Decision-Directed Tracking and Fusion}
The second refinement consists in using the fused soft symbol estimate \(\widehat c_{f,t}=\sum_{c \in \Cc} \bar p_{f,t}(c)\,c\) as a virtual pilot to form a decision-directed estimate,
\begin{equation}
    \widehat H^{\mathrm{ddtf}}_{f,t}=\frac{Y_{f,t}\,\widehat c^{\,*}_{f,t}}{\widehat{E}_{f,t}}, \qquad \widehat{E}_{f,t}=\sum_{c \in \Cc} \bar p_{f,t}(c)\,\abs{c}^2.
    \label{eq:gen_dd}
\end{equation}

Each \gls{re} is assigned a reliability
\begin{equation}
    \label{eq:gen_dd_rel}
    \eta_{f,t} = \chi_{f,t}\,\frac{|H^{\star}_{f,t}|^2}{|H^{\star}_{f,t}|^2+\sigma^2_{\mathrm{ddtf},f,t}}
\end{equation}
with
\begin{equation}
    \sigma^2_{\mathrm{ddtf},f,t} = \frac{N_0+|H^{\star}_{f,t}|^2\,\mathrm{Var}_{f,t}(c)}{\widehat{E}_{f,t}},
\end{equation}
where \(H^{\star}_{f,t}=\sum_k \pi_{f,t}(k)\,\widehat H^{(k)}_{f,t}\) is the trust-weighted mean channel, \(\mathrm{Var}_{f,t}(c)=\widehat{E}_{f,t}-\abs{\widehat c_{f,t}}^2\) is the soft-symbol variance, and \(\chi_{f,t}\in[0,1]\) is a symbol-confidence factor whose computation details are omitted.
The estimate is smoothed by a reliability-weighted Gaussian convolution,
\begin{equation}
    \widehat H^{\mathrm{ddtf}}_{f,t}\leftarrow\frac{\sum_{f',t'} \Gc_{f-f',\,t-t'}\,\eta_{f',t'}\,\widehat H^{\mathrm{ddtf}}_{f',t'}}{\sum_{f',t'} \Gc_{f-f',\,t-t'}\,\eta_{f',t'}},
    \label{eq:gen_dd_smooth}
\end{equation}
with \(\Gc\) a fixed Gaussian kernel, so that confidently estimated \glspl{re} dominate and unreliable ones are filled in from their neighbors.
The same convolution returns a channel-estimation variance \(\sigma^2_{H,f,t}\) from the smoothed \(\sigma^2_{\mathrm{ddtf}}\).

The smoothed track is appended to the pool as an additional candidate, with its own variance \(\sigma^2_{H,f,t}\).
All \(K+1\) candidates are then re-scored to update the trust weights \(\pi_{f,t}(k)\) by the same construction as~\eqref{eq:gen_trust}.
Each candidate is rated by its local mixture log-likelihood of the form~\eqref{eq:gen_score}, except that the noise is inflated by the channel estimation variance, i.e., \(N_0+\abs{c}^2\sigma^2_{H,f,t}\) is used instead of \(N_0\).

\subsection{Local Affine Correction}
The third refinement denoises the channel within each small time--frequency patch by fitting an affine surface
\begin{equation}
    H(f+\delta f,\,t+\delta t)\approx p_0 + p_1\delta t + p_2\delta f=\mathbf d^\top\mathbf p
    \label{eq:gen_affine_model}
\end{equation}
over a \(5\times3\) window \(\delta t\in\{-2,\dots,2\}\), \(\delta f\in\{-1,0,1\}\) centered on \((f,t)\), where \(\mathbf d=[1,\delta t,\delta f]^\top\) and \(\mathbf p=[p_0,p_1,p_2]^\top\in\CC^3\), so that \(p_0\) is the channel at the patch center and \(p_1,p_2\) are its temporal and spectral gradients.
The parameters minimize the following objective over the patch,
\begin{multline}
    J(\mathbf p)=\sum_{\delta f,\delta t} w_{f',t'}\,\EE\big[\,\abs{Y_{f',t'}-(\mathbf d^\top\mathbf p)\,c_{f',t'}}^2\,\big]\\
    +\lambda_0\abs{p_0-H^{\star}_{f,t}}^2+\lambda_s\LB\abs{p_1}^2+\abs{p_2}^2\RB,
    \label{eq:gen_affine_obj}
\end{multline}
where \((f',t')=(f+\delta f,t+\delta t)\) and the expectation is taken over the fused soft symbol at each patch \gls{re}, yielding \(\EE[\,\abs{Y-(\mathbf d^\top\mathbf p)c}^2\,]=\abs{Y-(\mathbf d^\top\mathbf p)\widehat c}^2+\abs{\mathbf d^\top\mathbf p}^2\,\mathrm{Var}(c)\).
The weight \(w_{f',t'}=\max\!\LB\max_{c\in\Cc} \bar p_{f',t'}(c),\,w_0\RB^{a}\) grows with the demapping confidence of the \gls{re}, so that reliably demapped \glspl{re} contribute most to the fit.
The regularization anchors \(p_0\) to the trust-weighted prior \(H^{\star}_{f,t}=\sum_k \pi_{f,t}(k)\,\widehat H^{(k)}_{f,t}\) with strength \(\lambda_0\) and shrinks the slopes with strength \(\lambda_s\).
Being quadratic in \(\mathbf p\), \eqref{eq:gen_affine_obj} is solved in closed form,
\begin{multline}
    \mathbf p=\Big(\textstyle\sum_{\delta f,\delta t} w_{f',t'}\,\widehat{E}_{f',t'}\,\mathbf d\mathbf d^\top+\mathbf R\Big)^{-1}\cdot\\
    \Big(\textstyle\sum_{\delta f,\delta t} w_{f',t'}\,\widehat c^{\,*}_{f',t'}\,Y_{f',t'}\,\mathbf d+\mathbf r_0\Big),
    \label{eq:gen_affine_sol}
\end{multline}
with \(\mathbf R=\mathrm{diag}(\lambda_0,\lambda_s,\lambda_s)\) and \(\mathbf r_0=[\lambda_0 H^{\star}_{f,t},0,0]^\top\). Here, \(\widehat{E}=\abs{\widehat c}^2+\mathrm{Var}(c)\).
The center value \(p_0\) is blended with its \(3\times3\) neighborhood for spatial consistency, then replaces the dominant candidate through a soft gate that opens according to a confidence threshold.
The channel-estimation variance \(\sigma^2_{H,f,t}\) is correspondingly inflated by the patch residual and the magnitude of the applied correction.

\subsection{Demapper}
Finally, the bit \glspl{llr} are computed by a demapper that sums the likelihood over both the constellation points and the competing channel candidates.
It first forms a per-symbol log-metric in which the noise variance of each candidate is inflated by its own channel-estimation variance.
The \(K\) candidates refined by the \gls{em} loop share the common variance \(\sigma^2_{H,f,t}\), whereas the appended decision-directed track carries its own variance and the affine correction inflates the variance of the candidate it replaces. Together these define the per-candidate variance \(\sigma^{2,(k)}_{H,f,t}\).
Let \(\nu^{(k)}_{f,t}(c)\triangleq N_0+\abs{c}^2\sigma^{2,(k)}_{H,f,t}\) be the inflated noise variance. Then
\begin{multline}
    \Lambda_{f,t}(c)= \ln{\sum_{k}\frac{\pi_{f,t}(k)}{\nu^{(k)}_{f,t}(c)}\,\exp{-\frac{\abs{Y_{f,t}-\widehat H^{(k)}_{f,t}\,c}^2}{\nu^{(k)}_{f,t}(c)}}}\\
    + w_\Gamma\,\Gamma_{f,t}(c).
    \label{eq:gen_metric}
\end{multline}
The ratio refinement \(\Gamma_{f,t}\) is added as a log-prior of weight \(w_\Gamma\).
Because the inflated variance also depends on the symbol through \(\abs{c}^2\), this normalizer is symbol-dependent and penalizes the higher-energy symbols.
The raw \gls{llr} of bit \(b\) marginalizes this metric over the subsets \(\Cc_{b,1}\) and \(\Cc_{b,0}\) of constellation points whose \(b\)th bit equals \(1\) and \(0\), respectively,
\begin{equation}
    \mathrm{LLR}^{\mathrm{raw}}_{f,t,b} = \ln{\frac{\sum_{c\in\Cc_{b,1}} e^{\Lambda_{f,t}(c)}}{\sum_{c\in\Cc_{b,0}} e^{\Lambda_{f,t}(c)}}}.
    \label{eq:gen_llr}
\end{equation}
Since \(\Gamma_{f,t}\) has already been injected upstream into the channel estimates and trust weights, and through~\eqref{eq:gen_metric} into the metric itself, the raw \gls{llr} is not extrinsic to the prior.
To provide an approximation of the demapper's extrinsic information, the receiver subtracts a reliability-weighted fraction of the prior's standalone bit-\gls{llr} \(L^{\Gamma}_{f,t,b}\),
\begin{equation}
    \mathrm{LLR}_{f,t,b} = \mathrm{LLR}^{\mathrm{raw}}_{f,t,b} - \lambda_{f,t}\,L^{\Gamma}_{f,t,b},
    \label{eq:gen_llr_damp}
\end{equation}
where the damping \(\lambda_{f,t}\) grows where \(\Gamma_{f,t}\) is less reliable, so an uncertain prior is suppressed more strongly in the output.

\section*{Acknowledgment}
The candidate algorithm code discussed in Sections~\ref{sec:otfs-equalizer-design}--\ref{sec:pilotless} and Appendixes~\ref{sec:otfs-candidate-description}--\ref{sec:pilotless-candidate-description} was generated by GPT-5.5 within \gls{aite}; the authors verified the implementations and edited the accompanying descriptions. OpenAI Codex with GPT-5.5 was used to check for grammar and spelling errors.

\bibliographystyle{IEEEtran}
\bstctlcite{IEEEexample:BSTcontrol}
\IEEEtriggeratref{45}
\bibliography{IEEEabrv,bibliography}

\end{document}